\documentstyle[w-edbkps,psfig]{w-edbk}
\newcommand{\I}{{\rm i}}
\newcommand{\e}{{\rm e}}

\newcommand{\s}{{\!_{(s)}}}
\newcommand{\norm}{{\cal N}_s}
\newcommand{\balpha}{{\overline\alpha\,}}
\newcommand{\bzeta}{{\overline\zeta\,}}
\newcommand{\bbeta}{{\overline\beta\,}}
\def\Langle#1{{_{(#1)}\!\langle}}
\def\Rangle#1{{\rangle\!_{(#1)}}}
\def\fra#1#2{{\textstyle\frac{#1}{#2}\,}}
\def\Hilbert#1{{${\cal H}^{(#1)}$}}
\def\binom#1#2{{#1\choose#2}}
\newcommand{\CS}  {{$|\alpha\rangle\s$}}

\begin{document}

\tableofcontents

\title[QUANTUM-OPTICAL STATES IN FINITE-DIMENSIONAL
HILBERT SPACE I.~GENERAL FORMALISM]{}

\markboth{}{Quantum-optical states in FD Hilbert space. I General
formalism}

%% Please supply author names in upper and lower case within square
%% brackets and in uppercase in curly brackets, i.e.,
%% \author[The Author]{THE AUTHOR\footnote{Presently on leave at
%% NASA, Houston, Texas, USA.}}

\author[A Miranowicz, W Leo\'nski, and N Imoto]{}

\begin{center}

{\bf  QUANTUM-OPTICAL STATES IN\\ FINITE-DIMENSIONAL HILBERT
SPACE.\\ I.~GENERAL FORMALISM}\footnote{Published in: {\em Modern
Nonlinear Optics, Part 1, Second Edition, Advances in Chemical
Physics, Vol. 119}, Edited by Myron W. Evans, Series Editors I.
Prigogine and Stuart A. Rice, 2001, John Wiley \& Sons, New York,
pp. 155--193.}

\vspace{4mm}%
ADAM MIRANOWICZ

\vspace{2mm}%
{\em CREST Research Team for Interacting Carrier
Electronics, School of Advanced Sciences, The Graduate University
for Advanced Studies (SOKEN), Hayama, Kanagawa, Japan and
Nonlinear Optics Division, Institute of Physics, Adam Mickiewicz
University, Pozna\'n, Poland}

\vspace{3mm}%
WIES\L{}AW LEO\'NSKI

\vspace{3mm}%
{\em Nonlinear Optics Division, Institute of Physics, Adam
Mickiewicz University, Pozna\'n, Poland}

\vspace{3mm}%
NOBUYUKI IMOTO

\vspace{2mm}%
{\em CREST Research Team for Interacting Carrier Electronics,
School of Advanced Sciences, The Graduate University for Advanced
Studies (SOKEN), Hayama, Kanagawa, Japan}

\end{center}

%%%%%%%%%%%%%%%%%%%%%%%%%%%%%%%%%%%%%%%%%%%%%%%%%%%%%%%%%%%%%%%%%%%%%%%%%%
\section{I.~~~~ INTRODUCTION}
\label{sect1}
\inxx{finite-dimensional Hilbert space}

In the late twentieth century much attention has been paid to the
investigation of various quantum-optical states defined in a {\em
finite-dimensional Hilbert space} of operators, which are bounded
and have a discrete spectrum. Yet, the idea of creating
finite-dimensional quantum-optical states was conceived much
earlier. In fact, back in 1931, Weyl's formulation of quantum
mechanics \cite{Wey31} opened the possibility of studying the
dynamics of quantum systems both in infinite-dimensional (ID) and
finite-dimensional (FD) Hilbert spaces. Weyl's approach,
generalized by Schwinger \cite{Sch60}, is based on the fact that
the kinematical structure of a physical system can be expressed
by an irreducible Abelian group of unitary representations of
system space. For a given finite Abelian group there is a unique
class of unitarily equivalent, irreducible representations in FD
space.  Hence, this formulation has provided the basis for
studies of the behavior of the harmonic oscillator in FD Hilbert
spaces. In the 1970s, Santhanam and coworkers contributed to the
above-mentioned formulation in a series of papers \cite{San76}. To
describe the interaction of an assembly of two-level atoms with a
transverse electromagnetic field, Radcliffe \cite{Rad71} and
Arecchi et al. \cite{Are72} introduced the atomic (or spin)
coherent states (also referred to as the directed
angular-momentum states \cite{Gla76}) as FD analogs of the
conventional optical coherent states (CS)
\cite{Sch26,Gla63,Sud63}. General formulation  of coherent states
in FD and ID Hilbert spaces was then developed by Perelomov
\cite{Per72}, and Gilmore et al. \cite{Gil72,Zha90} (see also
Refs. \cite{Mal79,Kla85,Fen94}). In the 1990s, various
quantum-optical states were constructed in FD Hilbert spaces in
analogy to those in the ID spaces. In particular, (1) various
kinds of FD coherent states
\cite{Buz92}--%Kua93,Mir94,Pat95,Opa96,Mir97,Roy97a,Roy97b,
\cite{Roy98}, (2) FD displaced number states \cite{Mir97}, (3) FD
even and odd coherent states \cite{Zhu94,Mir97,Roy98}, (4) FD
phase states \cite{Peg88}, (5) FD phase coherent states (also
referred to as coherent phase states)
\cite{Kua94}--%,Gan94,Mir95,
\cite{Roy97c}, (6) FD squeezed states \cite{Wod87}--%Fig93,Win94,
\cite{Mir98}, (7) FD displaced phase states \cite{Gan94}, or (8)
FD even and odd phase coherent states \cite{Kua96}.

The interest in the FD quantum-optical states has been stimulated
by the progress in quantum-optical state preparation and
measurement techniques \cite{state}, in particular, by the
development of the discrete quantum-state tomography
\cite{Ulf95}--% Man97,Buz97,Wal96,Ami98
\cite{Wel99}. There are several other reasons for studying states
in FD spaces.
\begin{enumerate}
\item
We can treat FD quantum-optical states as those of a real
single-mode electromagnetic field, which fulfill the condition of
truncated Fock expansion. These states can {\em directly} be
generated by the truncation schemes (the {\em \inx{quantum
scissors}}) proposed by Pegg et al. \cite{Peg98} and then
generalized by other authors \cite{Kon00,Par00,Mir00}.
Alternatively, one can analyze states obtained by a {\em direct}
truncation of operators rather then of their Fock expansion. Such
an operator truncation scheme, proposed by Leo\'nski et al.
\cite{Leo94,Leo97,Mir96a}, will be discussed in detail in the
next chapter \cite{Leo01}.

\item
The formalism of FD quantum-optical states is applicable to other
systems described by the FD models as well, such as spin systems
or ensembles of two-level atoms or quantum dots. In such cases we
should talk about, for instance, the $z$-component of the spin
and its azimuthal orientation rather than about the photon number
and phase. However, the states studied here were first discussed
in the quantum-optical papers and we also will keep the
terminology of quantum optics.

\item
This analysis gives us a deeper insight into the
\inx{Pegg--Barnett phase formalism} \cite{Peg88} (for a review,
see Ref. \cite{Tan96}) of the Hermitian optical phase operator
constructed in ($s+1$)-dimensional state Hilbert space. The key
idea of the Pegg--Barnett procedure is to calculate all the
physical quantities such as expectation values or variances in
the FD space and only then to take the limit of $s\rightarrow
\infty$. Bu\v{z}ek et al. \cite{Buz92} pointed out that all
quantities (in particular states) analyzed within the
Pegg--Barnett formalism, should properly be defined in the same
($s+1$)-dimensional state space before finally going over into
the infinite limit. So, for better understanding of the
Pegg--Barnett formalism, it is useful to construct
finite-dimensional states and to know what exactly happens before
taking the limit.
\end{enumerate}

\inxx{Wigner function; discrete} In this chapter, we apply a
discrete Wigner function to describe  FD quantum-optical states.
Wigner function is widely used in nonrelativistic quantum
mechanics as an alternative to the density matrix of quantum
systems \cite{Wig32}. Although the original Wigner function
applies only to systems with continuous degrees of freedom, it can
be generalized for finite-state systems as well \cite{Str57}.
Discrete Wigner function for spin-$\frac 12$ systems was
introduced by O'Connell and Wigner \cite{Oco84} and generalized
for arbitrary spins by Wootters \cite{Woo87}. His definition takes
the simplest form for prime-number-dimensional systems. A similar
construction of a discrete Wigner function for odd-dimensional
systems was suggested by a Cohendet et al. \cite{Coh88}. A
number-phase discrete Wigner function, a special case of the
Wootters definition, was analyzed in detail by Vaccaro and Pegg
\cite{Vac90}. Another definition of Wigner function (for odd
dimensions equivalent to that of Wootters) was proposed by
Leonhardt \cite{Ulf95}. This approach can readily be generalized
to define a discrete Husimi $Q$ function or, moreover, discrete
parameterized phase-space functions as was studied by Opatrn\'y et
al. \cite{Opa95,Opa96a}. Another generalization of discrete Wigner
function for Schwinger's FD periodic Hilbert space was analyzed
by, for instance, Hakio\v{g}lu \cite{Hak98}. The Wigner function
approach to FD systems can be developed from basic principles as
was shown, for example, by Wootters \cite{Woo87}, Leonhardt
\cite{Ulf95}, Luk\v{s} and Pe\v{r}inov\'a \cite{Luk93}, or Luis
and Pe\v{r}ina \cite{Lui98}. Discrete Wigner function has
successfully been applied to quantum-state tomography of FD
systems \cite{Ulf95} (for a review, see Ref. \cite{Wel99}).

This work is intended as an attempt to present two essentially
different constructions of harmonic oscillator states in a FD
Hilbert space. We propose some new definitions of the states and
find their explicit forms in the Fock representation. For the
convenience of the reader, we also bring together several known FD
quantum-optical states, thus making our exposition more
self-contained. We shall discuss FD coherent states, FD phase
coherent states, FD displaced number states, FD Schr\"odinger
cats, and FD squeezed vacuum. We shall show some intriguing
properties of the states with the help of the discrete Wigner
function.

%%%%%%%%%%%%%%%%%%%%%%%%%%%%%%%%%%%%%%%%%%%%%%%%%%%%%%%%%%%%%%%%%%%%%%%%%%
\section{II.~~~ FD HILBERT SPACE}
\label{sectFDHS} \inxx{finite-dimensional Hilbert space}

We shall discuss various states constructed in FD Hilbert space
of harmonic oscillator. Let us denote by \Hilbert{s} the
($s+1$)-dimensional Hilbert space spanned by number states
$\{|0\rangle,|1\rangle,\dots,|s\rangle\}$ fulfilling the
completeness and orthogonality relations
%--------------------------------------------------------------------------
\begin{equation}
\hat{1}_s=\sum_{n=0}^s|n\rangle \langle n|\,,\qquad \langle
n|m\rangle =\delta _{n,m} \label{N01}
\end{equation}
where $n,m=0,\dots,s$ and $\hat{1}_s$ is the unit operator in
\Hilbert{s}. Thus, arbitrary quantum-optical pure state in the FD
Hilbert space can be defined by its Fock expansion
%----------------------------------------------------------------------
\begin{equation}
 |\psi \Rangle{s}  =\;\sum_{n=0}^s C_n^{(s)}
|n\rangle \equiv\;\sum_{n=0}^s b_n^{(s)}
 \e ^{\I \varphi _n}|n\rangle \label{N02}
\end{equation}
where $C_n^{(s)}=b_n^{(s)} \e^{\I \varphi _n}$ and $b_n^{(s)}$ are
real superposition coefficients fulfilling the normalization
condition
%----------------------------------------------------------------------
\begin{equation}
\Langle{s} \psi |\psi \Rangle{s} =\sum_{n=0}^s [b_{n}^{(s)}]^2=1
\label{N03}
\end{equation}
for arbitrary dimension $(s+1)$ of Hilbert space. It is sometimes
useful to represent the optical state, given by (\ref{N02}), via
the phase states defined to be \cite{Peg88} \inxx{phase states}
%--------------------------------------------------------------------------
\begin{equation}
 |\theta _m\rangle\equiv |\theta _m\Rangle{s} =\frac 1{\sqrt{s+1}}\sum_{n=0}^s\exp
(\I n\theta _m)|n\rangle \label{N04}
\end{equation}
with the phases $\theta _m$ given by
%--------------------------------------------------------------------------
\begin{equation}
\theta _m=\theta _0+\frac{2\pi}{s+1}\,m \label{N05}
\end{equation}
where $\theta _0$ is the initial reference phase and
$m=0,\dots,s$. States (\ref{N04}) also form a complete and
orthonormal basis:
%--------------------------------------------------------------------------
\begin{equation}
 \hat{1}_s=\sum_{m=0}^s|\theta _m\rangle \langle
\theta _m|\,,\qquad \langle \theta _m|\theta _n\rangle =\delta
_{m,n}\label{N06}
\end{equation}
The phase states were applied by Pegg and Barnett in their
definition of the Hermitian quantum-optical phase operator
\cite{Peg88} \inxx{phase operator}
%----------------------------------------------------------------------
\begin{eqnarray}
{\hat\Phi}_s &\equiv& {\hat\Phi}_s(\theta_0)\;=\; \sum_{m=1}^{s}
\theta_m|\theta_{m}\rangle\langle \theta_{m}| \label{N07}
\end{eqnarray}
The phase states can also be used in construction of a discrete
Wigner function as will be described in Section III. The FD
annihilation and creation operators in \Hilbert{s} are defined by
%---------------------------------------------------------------------------
\begin{eqnarray}
\hat{a}_s&=&\sum_{n=1}^s\sqrt{n} |n-1\rangle \langle n|
\nonumber \\
\hat{a}_s^{\dagger}&=&\sum_{n=1}^s\sqrt{n}|n\rangle \langle n-1|
\label{N08}
\end{eqnarray}
The FD and ID annihilation operators act on a number state in the
same manner. However, the actions of the creation operators on
$|n\rangle $ are different in \Hilbert{s} and \Hilbert{\infty}.
Equation (\ref{N08}) implies that
%----------------------------------------------------------------------
\begin{eqnarray}
({\hat{a}}_s^{\dagger})^k |n\rangle =0 \label{N09}
\end{eqnarray}
if $n+k>s$. By contrast, the action of the ID creation operator
(in any power) on $|n\rangle$ gives always nonzero result. The
commutation relation for the annihilation and creation operators
in \Hilbert{s} reads as
%----------------------------------------------------------------------
\begin{equation}
[{\hat{a}_s},{\hat{a}_s^{\dagger }]=1-(s+1)|s\rangle \langle s|}
\label{N10}
\end{equation}
which differs from the conventional boson canonical relation in
\Hilbert{\infty}. Thus, $\hat{a}_s$ and $\hat{a}_s^{\dagger}$ are
not related to the Weyl--Heisenberg algebra. Even the double
commutators $[\hat{a}_s,[\hat{a}_s,\hat{a}_s^{\dagger}]]$ and
$[\hat{a}_s^{\dagger},[\hat{a}_s,\hat{a}_s^{\dagger }]]$ do not
vanish precluding the application of the Baker-Hausdorff theorem.
These properties of the FD annihilation and creation operators
considerably complicate analytical approaches to the quantum
mechanics in \Hilbert{s}, including the explicit construction of
the FD harmonic oscillator states.

Creation and annihilation of phase quanta in FD Hilbert space can
be defined in a close analogy to the creation and annihilation of
photons, as given by Eq. (\ref{N08}). Phase annihilation,
$\hat\phi_s$, and phase creation, $\hat\phi_s^{\dagger}$,
operators can be introduced with the help of the relation
$\hat\Phi_s = \hat\phi_s ^{\dagger} \hat\phi_s$ for the
Pegg--Barnett phase operator, given by (\ref{N07}). The FD phase
annihilation and creation operators in the phase-state basis,
have the following form \cite{Buz92}
%----------------------------------------------------------------------
\begin{eqnarray}
\hat\phi_s\;\equiv\; \hat\phi_s(\theta_0) &=& \sum_{m=1}^{s}
\sqrt{\theta_m}|\theta_{m-1}\rangle\langle
\theta_{m}|+\sqrt{\theta_0}|\theta_{s}\rangle\langle\theta_{0}|
\nonumber\\
\hat\phi_s^{\dagger}\;\equiv\; \hat\phi_s^{\dagger}(\theta_0) &=&
\sum_{m=1}^{s} \sqrt{\theta_m}|\theta_{m}\rangle\langle
\theta_{m-1}|
+\sqrt{\theta_0}|\theta_{0}\rangle\langle\theta_{s}| \label{N11}
\end{eqnarray}
respectively. Their commutator is
%----------------------------------------------------------------------
\begin{eqnarray}
[\hat\phi_s,\hat\phi_s^{\dagger}] &=&
\frac{2\pi}{s+1}-2\pi|\theta_{s}\rangle\langle \theta_{s}|
\label{N12}
\end{eqnarray}
The phase annihilation and creation operators act on the phase
states in a similar way (particularly for $\theta_0=0$) as the
conventional (photon-number) annihilation and creation operators
act on number states.

%%%%%%%%%%%%%%%%%%%%%%%%%%%%%%%%%%%%%%%%%%%%%%%%%%%%%%%%%%%%%%%%%%%%%%%%%%
\section{III.~\ DISCRETE WIGNER FUNCTION FOR FD STATES}
\label{sectWigner} \inxx{Wigner function; discrete}

The expression of quantum-optical states by quasidistributions
enables a very intuitive description of their properties. Here, we
give a general definition of the discrete Wigner-function. We also
present its graphical representations for several FD
quantum-optical states, often discussed in recent works. The
Wigner function will be studied in greater detail in the following
sections.

The number--phase \inx{characteristic function} in \Hilbert{s} can
be defined as \cite{Ulf95}
%---------------------------------------------------------------------------
\begin{equation}
 C_s(\nu, \theta _\mu )=\sum_{m=0}^s\exp \left(
-\frac{4\pi \I }{s+1} \nu (m+\mu )\right) \langle \theta
_m|\hat{\rho}|\theta _{m+2\mu }\rangle  \label{N13}
\end{equation}
in terms of the phase states (\ref{N04}). A discrete Fourier
transform applied to $C_s(\nu, \theta _\mu )$ leads to the
following discrete Wigner function (for brevity referred to as the
$W$ function) for phase and number
%---------------------------------------------------------------------------
\begin{equation}
 W_s(n,\theta _m)=\frac 1{\left( s+1\right) ^2}\sum_{\nu
=0}^s\sum_{\mu =0}^s\exp \left(\frac{4\pi \I }{s+1}(n\mu +\nu
m)\right) C_s(\nu ,\theta _\mu ) \label{N14}
\end{equation}
%%%%%%%%%%%%%%%%%%%%%%%%%%%%%%%%%%%%%%%%%%%%%%%%%%%%%%%%%%%%%%%%%%%%%%%
% figure 1
\begin{figure}
\vspace*{-4.5cm} \hspace*{1cm}
\centerline{\psfig{figure=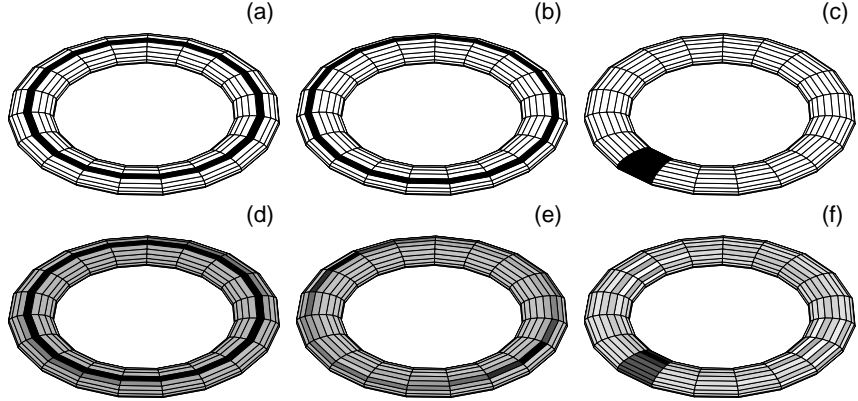,width=17cm}}
\vspace*{-15cm} \caption{Examples of discrete Wigner function on a
torus in 19-dimensional Hilbert space ($s=18$):  {(a)} vacuum
$|0\rangle$; {(b)} single-photon number state $|1\rangle$; {(c)}
FD preferred phase state (``phase vacuum'')
$|\theta_0\rangle_{(s)}$; {(d)} FD coherent state,
$|\alpha\rangle_{(s)}\approx |\overline{\alpha}\rangle_{(s)}$;
{(e)} FD displaced number state, $|\alpha,1\rangle_{(s)}\approx
|\overline{\alpha}, 1\rangle_{(s)}$; {(f)} FD phase coherent
state,
$|\beta,\theta_0\rangle_{(s)}\approx|\overline{\beta},\theta_0
\rangle_{(s)}$, with equal displacement parameters, $\alpha=
\overline{\alpha}= \beta= \overline{\beta}=0.5$ and $\theta_0=0$.
The darker is a region, the higher is the value of the Wigner
function. \label{mirafg01}} \vspace*{-0.5cm}
\end{figure} \noindent%
or, explicitly, as \cite{Woo87,Ulf95}
%---------------------------------------------------------------------------
\begin{equation}
W_s(n,\theta _m)=\frac 1{s+1}\sum_{\mu =0}^s\exp \left(\frac{
4\pi \I }{s+1}n\mu \right) \langle \theta
_{m-\mu}|\hat{\rho}|\theta _{m+\mu }\rangle \label{N15}
\end{equation}
The Wigner function $W_s(n,\theta _m)$ is periodic both in $n$
and $\theta_m$:
%---------------------------------------------------------------------------
\begin{eqnarray}
W_s(n,\theta _m)&=&W_s(n\pm \{s+1\},\theta _m) \nonumber\\
&=&W_s(n,\theta _{m\pm (s+1)})\nonumber\\ &=& W_s(n,\theta _m\pm
2\pi ) \label{N16}
\end{eqnarray}
Thus, it is represented graphically on torus \cite{Opa96}.  The
Wigner function for any FD pure state of the form (\ref{N02}) can
be expressed as follows \cite{Vac90}
%----------------------------------------------------------------------
\begin{eqnarray}
 W_s(n, \theta _{m}) = \frac 1{s+1} \left\{ \sum_{k=0}^{M}
b^{(s)}_{k} b^{(s)}_{M-k} \exp[\I  (2k - M) \theta
_{m}+\varphi_{M-k}-\varphi_{k}] \right. \quad
\nonumber \\
\;+ \!\!\!\!\left. \sum_{k=M+1}^{s} b^{(s)}_{k} b^{(s)}_{M-k+s+1}
\exp [\I  (2k - M -s -1) \theta
_{m}+\varphi_{M-k+s+1}-\varphi_{k}] \right\} \label{N17}
\end{eqnarray}
in terms of the decomposition coefficients $b^{(s)}_{k}$ and
$M\equiv 2n\ \mbox{mod}(s+1)$. Several graphs of these functions
of various states are presented here (Fig. \ref{mirafg01}) and in
the next sections. The physical interpretation of the Wigner
functions is based on the fact that the marginal sum of their
values over a generalized line gives the probability that the
system will be in some state \cite{Woo87,Ulf95}. In the ID
Hilbert space, where the $W$ function arguments are continuous
(quadratures $X$ and $Y$), a marginal integral along any straight
line $aX + bY + c = 0$ is nonnegative and can be considered to be
the probability. A similar situation arises in the FD case; we can
define lines as sets of discrete points $(n, \theta _{m})$, or
equivalently $(n,m)$, for which the relation $(an + bm +
c)\mbox{mod}N = 0$ holds (here, $a, b, c$ are integers). Again,
sums of the discrete $W$ function values on such sets are
nonnegative. The mod $(s+1)$ relations are essential and are
connected to some periodic properties of the discrete $W$ function
--- the maximum value of each argument ($m$ or $n$) is
topologically followed by its minimum (zero in our case). This
means that the discrete $W$ function is defined on a torus (or
more precisely on a discrete set of points of a torus). The
``lines'' are then points of closed toroidal spirals or, in a
special case, points of a circle. The periodic property is quite
natural for the phase index $m$, but may seem strange for the
photon number $n$. In the next sections, we shall draw attention
to some consequences of the periodicity in $n$ for generalized
coherent states, for instance.

One aim of this chapter is to show graphs of the discrete $W$
functions for FD quantum-optical states. Because of the
discreteness of the arguments, the $W$ function graph should be a
histogram. However, two-dimensional projections of such
three-dimensional histograms could be very confusing. Therefore,
for better legibility of the graphs, we have decided to depict
them topographically. The darker is a region, the higher is the
value of the $W$ function it represents. Moreover, negative
values of the $W$ function are marked by crosses. As mentioned
above, the most natural way of presenting the discrete $W$
function graphs is to construct them on toruses. A few simple
examples of the toroidal discrete Wigner functions are given in
Fig. \ref{mirafg01}. Unfortunately, this graphical representation
is seldom transparent enough for its interpretation. In what
follows we shall work with two-dimensional graphs. Here, one
should keep in mind that some consequences of the periodicity in
$n$ and $m$ can appear: for instance, some peaks can be located
partially at the outer boundary at $n \approx s$ (or $m \approx
s$) and can ``continue'' near the center $n \approx 0$ (or $m
\approx 0$). In the next section, the $W$ functions of various FD
states will be presented.

For a better understanding of the discrete Wigner function, let us
recall its close correspondence to the discrete \inx{Pegg--Barnett
phase distribution} \cite{Peg88}
%----------------------------------------------------------------------
\begin{eqnarray}\label{N18}
P_s(\theta_{m}) = \sum_{n=0}^{s} W_s(n,\theta_{m})
=\left|\Langle{s}\theta_{m}|\psi\Rangle{s}\right|^2
\end{eqnarray}
and to the \inx{photon-number distribution}
%----------------------------------------------------------------------
\begin{eqnarray}\label{N19}
P_s(n) &=& \sum_{m=0}^{s} W_s(n,\theta_{m})= \left|\langle
n|\psi\Rangle{s}\right|^2 =|b^{(s)}_n|^2
\end{eqnarray}
Thus, the sum of the $W$ function values, at constant $\theta
_{m}$, over all $n$ values gives the probability of the phase
$\theta _{m}$ and, analogously, the sum with constant $n$ over all
arguments $\theta _{m}$ gives the probability of $n$ photons ---
at least in systems, which are fully described by finite-number
state models. If we want to interpret our results as describing
states of a usual one-mode field under the condition that all Fock
components of $|n \rangle$ with $n > s$ are absent, then the real
phase probability distribution is obviously continuous. Let us
briefly discuss its connection to the obtained discrete
distribution. If $s$ is greater than or equal to the largest Fock
state component of a given state, which by definition is our case,
then the discrete probabilities (from the discrete Wigner phase
marginal) are proportional to the values of the continuous phase
probability distribution in the discrete set of points
(\ref{N05}). One can easily obtain other values also, although not
directly. We could use a FD version of the sampling theorem --- if
the $n$ distribution is limited, then for description a state in
the phase representation only a discrete set of phase amplitudes
is necessary. It is clear that the $(s+1)^{2}$ real values of the
discrete $W$ function yield the same information as the
$(s+1)^{2}$ real nonzero parameters of the related density matrix.

%%%%%%%%%%%%%%%%%%%%%%%%%%%%%%%%%%%%%%%%%%%%%%%%%%%%%%%%%%%%%%%%%%%%%%%%
\section{IV.~~ FD COHERENT STATES}
\label{sect4}

The most common states in quantum optics are the {\em coherent
states} (CS) introduced by Schr\"odinger \cite{Sch26} in
connection with classical states of the quantum harmonic
oscillator. First modern description and specific application of
CS is due to Glauber \cite{Gla63} and Sudarshan \cite{Sud63}. The
literature on CS and their generalizations is truly prodigious and
has been summarized in a number of excellent monographs
\cite{Per72,Mal79,Kla85,Per94} and reviews \cite{Zha90}. There are
several ways of generalizing the conventional ID coherent states
to comprise the FD case. It is possible to define CS using the
concept of Lie group representations \cite{Per94}, or to
postulate the validity of some properties of the ID CS for their
FD analogs. In this chapter, we are interested in two definitions
of the latter kind. First, CS in a FD space are usually treated as
the displaced vacuum, where the displacement operator is defined
analogously to the conventional displacement operator in the ID
space (the Glauber treatment of CS \cite{Gla63}). This idea was
applied in the work of Bu\v{z}ek and co-workers \cite{Buz92} and
further studied by Miranowicz et al. \cite {Mir94} and Opatrn\'y
et al. \cite{Opa96}. Here, we refer to such states as the {\sl
generalized CS}. Another definition is based on the postulate
that the Fock expansion of the FD CS is be equal to the truncated
expansion of the conventional ID CS. This approach was
extensively developed by Kuang et al. \cite{Kua93} and Opatrn\'y
et al. \cite{Opa96}. Here, we shall refer to CS of this kind as
the {\sl truncated CS}. An experimental scheme, known as the {\em
quantum scissors}, for generation of the truncated CS was
proposed by Pegg et al. \cite{Peg98}. Quantum scissors were
generalized by Koniorczyk et al. \cite{Kon00}, Paris \cite{Par00},
and Miranowicz et al. \cite{Mir00}. A physical system for
preparation of the generalized CS was proposed by Leo\'nski
\cite{Leo97} (see also Ref. \cite{Mir96a}) as a modification of
the Fock state engineering technique of Leo\'nski and Tana\'s
\cite{Leo94}. These schemes are presented in the next chapter
\cite{Leo01}.

%%%%%%%%%%%%%%%%%%%%%%%%%%%%%%%%%%%%%%%%%%%%%%%%%%%%%%%%%%%%%%%%%%%%%%%%
\subsection{A.~~ Generalized Coherent States}
\label{sectGCS} \inxx{coherent states; generalized}

Glauber  \cite{Gla63} constructed coherent states in the ID
Hilbert space by applying the displacement operator
$\hat{D}(\alpha,\alpha^*)\equiv \exp(\alpha\hat{a}^{\dagger}
-\alpha^* \hat{a})$ on vacuum state $|0\rangle$. Analogously, one
can define the {\em generalized coherent state} \cite{Buz92}
%----------------------------------------------------------------------
\begin{eqnarray} |\alpha\Rangle{s}  &=& \hat{D}_s
(\alpha,\alpha^*) |0\rangle \label{N20}
\end{eqnarray}
%%%%%%%%%%%%%%%%%%%%%%%%%%%%%%%%%%%%%%%%%%%%%%%%%%%%%%%%%%%%%%%%%%%%%%%
% figure 2
\begin{figure}
\vspace*{-4cm} \hspace*{3mm}
\centerline{\psfig{figure=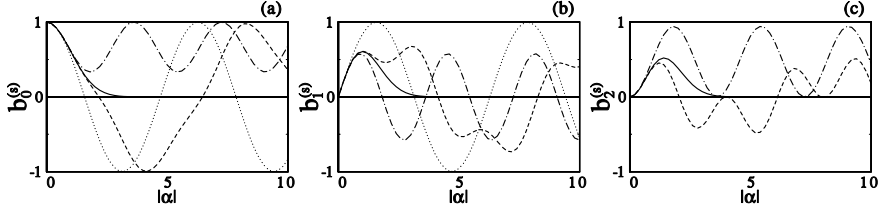,width=15cm}}
\vspace*{-15cm} \caption{{\bf Generalized coherent states}: The
superposition coefficients $b^{(s)}_{n}$ for
$|\alpha\rangle_{(s)}$ versus displacement parameter amplitude
 $|\alpha|$ for: (a) $n$=0, (b) $n$=1, and (c) $n$=2 in
the Hilbert spaces of different dimensionality:  $s=1$ (dotted),
$s=2$ (dot--dashed), $s=3$ (dashed), and $s=\infty$ (solid
curves). \label{mirafg02}}
%\vspace*{-1cm}
\end{figure}
%%%%%%%%%%%%%%%%%%%%%%%%%%%%%%%%%%%%%%%%%%%%%%%%%%%%%%%%%%%%%%%%%%%%%%%
% figure 3
\begin{figure}
\vspace*{-4.5cm} \hspace*{7mm}
\centerline{\psfig{figure=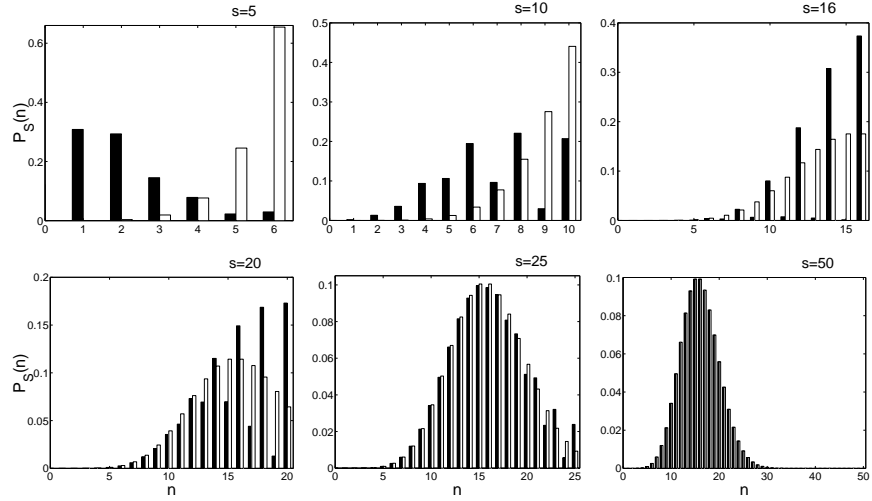,width=17cm}}
\vspace*{-13.8cm} \caption{{\bf Generalized coherent states}
(black bars) versus {\bf truncated coherent states} (white bars):
photon-number distribution $P_s(n)$ as a function of $n$ in FD
Hilbert spaces with $s=5,\cdots,50$ for the same displacement
parameters $\alpha=\overline\alpha=4$. \label{mirafg03}}
\vspace*{-0.7cm}
\end{figure} \noindent%
constructed in the FD Hilbert space by the action of the
generalized FD given by \inx{displacement operator}
\begin{eqnarray}
\hat{D}_s(\alpha,\alpha^*) &=& \exp\big[\alpha\hat{a}_s^{\dagger}
-\alpha^* \hat{a}_s\big] \label{N21}
\end{eqnarray}
where the FD annihilation and creation operators are given by
(\ref{N08}). Definitions of CS based on displacement operators are
usually applied in various generalizations of CS
\cite{Rad71,Are72,Per72,Zha90,Buz92,Mir94}. The generalized
coherent state, $|\alpha\Rangle{s}$ with $\alpha=|\alpha|\exp(\I
\varphi)$, has the following Fock expansion \cite{Mir94}
%----------------------------------------------------------------------
\begin{eqnarray}
 |\alpha\Rangle{s}  =\sum_{n=0}^{s} \e ^{\I  n\varphi} \,
 b^{(s)}_n \,|n\rangle \label{N22}
\end{eqnarray}
where
%----------------------------------------------------------------------
\begin{eqnarray}
 b^{(s)}_n &=& \frac{s!}{s+1} \frac{(-\I )^{n}}{\sqrt{n!}}\,  \sum_{k=0}^{s}
 \e ^{\I  x_k|\alpha|}\, \frac{{\rm
He}_n(x_k)}{{\rm He}_{s}^{2}(x_k)} \label{N23}
\end{eqnarray}
Here, $x_{k} \equiv x_{k}^{(s+1)}$ are the roots,
$\mbox{He}_{s+1}(x_{k}) = 0$, of the \inx{Hermite polynomial}
$\mbox{He}_{n}(x) \equiv 2^{-n/2} \mbox{H}_{n}(x/\sqrt{2})$. A
method for deriving the coefficients (\ref{N23}) is presented in
the Appendix.  In the special cases for $s=1,2,3$, the generalized
CS are as follows
%----------------------------------------------------------------------
\begin{eqnarray}
|\alpha\Rangle{1} &=&  \cos|\alpha| |0\rangle +\e ^{\I \varphi}
\sin|\alpha| |1\rangle \label{N24}
\end{eqnarray}
%----------------------------------------------------------------------
\begin{eqnarray}
|\alpha\Rangle{2} &=&
\frac{1}{3}\Big[\cos\big(\sqrt{3}|\alpha|\big)+2 \Big] |0\rangle
+\frac{1}{\sqrt{3}} \,\e ^{\I  \varphi}
\sin\big(\sqrt{3}|\alpha|\big) |1\rangle \nonumber\\ &&+
\frac{\sqrt{2}}{3} \,\e ^{2\I  \varphi}
\left[1-\cos\big(\sqrt{3}|\alpha|\big)\right] |2\rangle
\label{N25}
\end{eqnarray}
%----------------------------------------------------------------------
\begin{eqnarray}
 |\alpha\Rangle{3} &=& \frac{x^2_2{\sf c}_1
+x^2_1{\sf c}_2}{2x^2_1x^2_2} |0\rangle + \frac{x_2 {\sf s}_1 +x_1
{\sf s}_2}{2x_1x_2}\,\e ^{\I \varphi} |1\rangle \nonumber\\
&&- \frac{{\sf c}_1-{\sf c}_2}{2\sqrt{3}} \,\e ^{2\I  \varphi}
|2\rangle - \frac{x_2 {\sf s}_1 -x_1 {\sf s}_2}{2x_1x_2}\, \e
^{3\I  \varphi} |3\rangle \label{N26}
\end{eqnarray}
where ${\sf s}_k=\sin(x^{(4)}_k|\alpha|)$ and ${\sf
c}_k=\cos(x^{(4)}_k |\alpha|)$ are functions of the roots
$x_{1,2}^{(4)} = \sqrt{3\pm \sqrt{6}}$.   The state (\ref{N24}) in
the two-dimensional Hilbert space will be studied in greater
detail in Section IV.C. The simplicity of (\ref{N24}) comes from
the fact that the only nonvanishing coefficients $d_{nk}^{(1)}$,
given by Eq. (\ref{N102}), are equal to unity.  In Fig.
\ref{mirafg02}, the coefficients $b_{n}^{(s)}$ are presented in
their dependence on the parameter $|\alpha|$ for $s = 1,2,3$ and
$s=\infty$. It is seen that the coefficients (\ref{N23}) are
periodic (for $s$=1, 2) or quasiperiodic (for higher $s$) in
$|\balpha|$. The generalized CS go over into the conventional CS
in the limit of $s\rightarrow\infty$. This conclusion can be drawn
by analyzing Fig. \ref{mirafg03}, where the photon-number
distribution $P_s(n)=|b_{n}^{(s)}|^2$ for the generalized CS is
presented for different values of $s$ and fixed $\alpha=4$. The
differences between $|\alpha\Rangle{s}$ and $|\alpha
\Rangle{\infty}$  vanish even for $s=50$ on the scale of Fig.
\ref{mirafg03}. In order to prove this property analytically, let
us expand the scalar product between $|\alpha\Rangle{s} $ and
$|\alpha\Rangle{\infty}$ in series of parameter $|\alpha|$. One
finds the following power series expansions \cite{Opa96}
%----------------------------------------------------------------------
\begin{eqnarray}
\Langle{\infty}\alpha|\alpha\Rangle{1} &=& 1 - \fra 14 |\alpha|^4
+ \fra 19 |\alpha|^6 - {\cal O}(|\alpha|^8) \nonumber\\
\Langle{\infty}\alpha|\alpha\Rangle{2} &=& 1 - \fra 1{12}
|\alpha|^6 + \fra{3}{64} |\alpha|^8 - {\cal O}(|\alpha|^{10})
\nonumber\\
\Langle{\infty}\alpha|\alpha\Rangle{3} &=& 1 - \fra {1}{48}
|\alpha|^8 + \fra{1}{75}|\alpha|^{10} - {\cal O}(|\alpha|^{12})
\label{N27}
\end{eqnarray}
for particular values of $s$. We see that, with increasing
dimension, the generalized CS approach the conventional CS as
%----------------------------------------------------------------------
\begin{eqnarray}
 \Langle{\infty}\alpha|\alpha\Rangle{s}  &=& 1 -
\frac{|\alpha|^{2(s+1)}}{2\,(s+1)!} + {\cal O}(|\alpha|^{2(s+2)})
\label{N28}
\end{eqnarray}
for $|\alpha|^2\ll s$.

\begin{figure}
\vspace*{-4.2cm} \hspace*{8mm}
\centerline{\psfig{figure=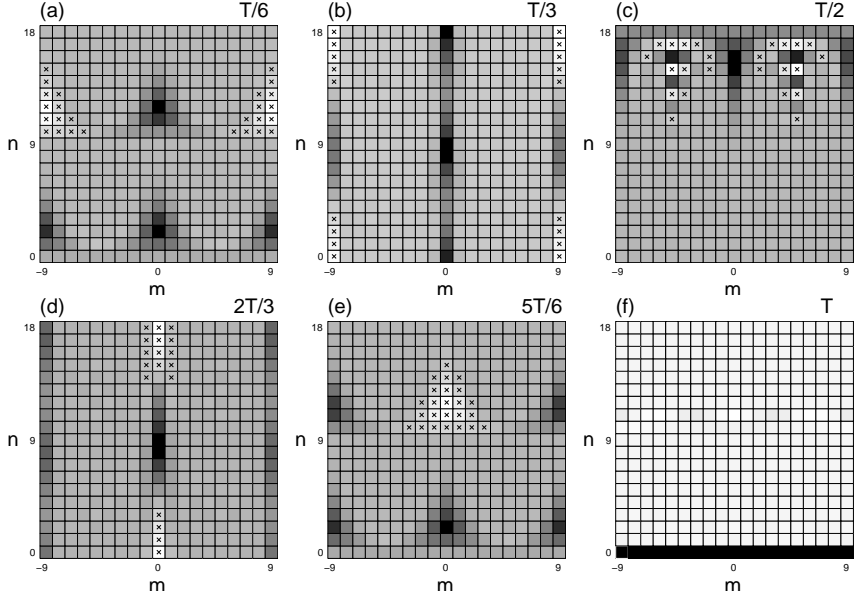,width=17cm}}
\vspace*{-12.5cm} \caption{{\bf Generalized coherent states}:
Wigner function $W_{s}(n,\theta_m)$ in FD Hilbert space with
$s=18$ for $|\alpha\rangle_{(18)}$ with different values of
displacement parameter $\alpha$, chosen as fractions of the
quasiperiod $T\equiv T_{18}\approx 8.8$. As in Fig.
\ref{mirafg01}, higher values of Wigner function are depicted
darker. Negative regions are marked additionally by
crosses.\label{mirafg04}} \vspace*{-0.6cm}
\end{figure}

On insertion of the coefficients (\ref{N23}) into the general
formula (\ref{N17}), we get the Wigner function for
$|\alpha\Rangle{s}$ in the form
%----------------------------------------------------------------------
\begin{eqnarray}
W_s(n, \theta _{m}) &=& \sum_{k=M+1}^{s} \frac{\exp [\I
(2k-M-s-1)(\theta _{m}-\varphi + \pi /2)]} {[k!(M-k+s+1)!]^{1/2}}
G_{1k}
\nonumber \\
&&+\! \sum_{k=0}^{M} \frac{\exp [\I  (2k-M)(\theta _{m}-\varphi +
\pi /2)]} {[k!(M-k)!]^{1/2}} G_{0k} \label{N29}
\end{eqnarray}
where
%----------------------------------------------------------------------
\begin{eqnarray}
\label{N30} G_{\eta k} = \frac{(s!)^{2}}{(s+1)^3} \sum_{p=0}^{s}
\sum_{q=0}^{s} \exp [\I  (x_{q}-x_{p})|\alpha |]
\frac{\mbox{He}_{k}(x_{p})\mbox{He}_{M-k+\eta(s+1)}(x_{q})}
{[\mbox{He}_{s}(x_{p})\mbox{He}_{s}(x_{q})]^{2}}
\end{eqnarray}
with $\eta=0,1$. By writing Eq. (\ref{N29}) in a form more similar
to the Vaccaro--Pegg expression, we arrive at
%----------------------------------------------------------------------
\begin{eqnarray}
\label{N31} W_s(n, \theta _{m}) &=& \sum_{k=2n+1}^{s}
(-1)^{k-n-s/2}\; \frac{\sin [(2k-2n-s-1)(\theta _{m}-\varphi)]}
{[k!(2n-k+s+1)!]^{1/2}} G_{1k}
\nonumber \\
&&+ \sum_{k=0}^{2n} (-1)^{k-n}\; \frac{\cos [(2k-2n)(\theta
_{m}-\varphi)]} {[k!(2n-k)!]^{1/2}} G_{0k}
\end{eqnarray}
for $n \le s/2$, and
%----------------------------------------------------------------------
\begin{eqnarray}
\label{N32} W_s(n, \theta _{m}) &=& \sum_{k=0}^{2n-s-1}
(-1)^{k-n-s/2}\; \frac{\sin [(2k-2n+s+1)(\theta _{m}-\varphi)]}
{[k!(2n-k-s-1)!]^{1/2}} G_{0k}
\nonumber \\
&&+\; \sum_{k=2n-s}^{s} (-1)^{k-n}\; \frac{\cos [(2k-2n)(\theta
_{m}-\varphi)]} {[k!(2n-k)!]^{1/2}} G_{1k}
\end{eqnarray}
for $n > s/2$. As readily seen, we cannot generally factorize this
function into a product of amplitude $|\alpha|$- and phase
$\varphi$- dependent parts. The Wigner functions for the
generalized CS are presented for $s=18$ in Fig. \ref{mirafg04}. We
observe the following behavior of the Wigner function. The shape
of the respective graph is approximately periodic (referred to as
the quasiperiodic) in the parameter $|\alpha|$ with quasiperiod
$T_{s} \approx 8.8$. We find that for small $|\balpha|$ the shape
is essentially the same as that described in Ref. \cite{Vac90} ---
for $n \le s/2$, there are two peaks for opposite phases, whereas
for $n > s/2$ we observe a peak and an antipeak. Note, the peaks
at the borders are artificially split up in our Cartesian
representation of the Wigner function. The peaks or antipeaks are
located at such positions that on summing the $W$ function with
constant $n$ (or $\theta _{m}$) over $\theta _{m}$ (or $n$), we
get the probability distribution of $n$ (or $\theta _{m}$,
respectively). Then, with increasing $|\alpha|$, interesting
oscillations in photon number appear. Their culmination is at
$|\alpha | = T_{s}/2$ (Fig. \ref{mirafg04}c), where only even
photon numbers are present. For this value of $\alpha$, the
generalized coherent state approaches an even CS, namely, the case
of a Schr\" odinger cat state, which will be described in detail
in Section IV.D.1. By further enlarging $|\alpha|$, the $W$
function returns to its previous shapes through the transition
regime (for $|\alpha | \approx 2T_{s}/3$ in Fig. \ref{mirafg04}d)
to the case of the inner two-peak and outer peak--antipeak
structure, similar to the Vaccaro--Pegg results. For $|\alpha |
\approx 5T_s/6$ as given in Fig. \ref{mirafg04}e, the $W$
function is very similar to that presented in Fig.
\ref{mirafg04}a for $|\alpha | \approx T_s/6$, but with opposite
phase. Finally, for $|\alpha | = T_s$ as presented in Fig.
\ref{mirafg04}f, we arrive at an almost vacuum state. By further
increasing $|\alpha |$, these shapes of the $W$ function graph
reappeared for several quasiperiods $T_s$. Similar behavior can
be observed also for other values of $s$.

This quasiperiodicity can be explained as follows. By applying
the fitting procedure, based on the WKB
(Wentzel--Kramers--Brillouin) method, one can find that the
smallest positive root $x_1\equiv x_1^{(s+1)}$ of the
\inx{Hermite polynomial} He$_{s+1}(x)$ is approximately equal to
%----------------------------------------------------------------------
\begin{equation}
 x_1^{(s+1)} \approx \frac{2\pi}{\sqrt{4s+6}} \label{N33}
\end{equation}
(for even $s$). Besides, it is well known that the
nearest-to-zero roots of the Hermite polynomials are
approximately equidistant. Thus, their difference $\Delta x
\equiv x_{k+1} - x_{k}$ is approximately given by (\ref{N33}),
which is 0.71 for $s=18$. The predominant terms of the sum in
(\ref{N23}) depend on $|\alpha|$ approximately as $\exp(\I
g\Delta x |\alpha |)$, where $g = 0, \pm 1, \pm 2, \dots$. These
exponential functions are quasiperiodic with approximate mean
period (referred to as the {\em quasiperiod}) given by
\cite{Opa96,Leo96}
%----------------------------------------------------------------------
\begin{eqnarray}
 T_s\approx \sqrt{4s+6} \label{N34}
\end{eqnarray}
for even $s$. From Eq. (\ref{N34}), the quasiperiod for $s$=18 is
approximately equal to $8.8$. For odd $n$, the property
$\mbox{He}_{n}(-x_{k}) = (-1)^{n}\mbox{He}_{n}(x_{k})$ holds. On
the other hand, the odd coefficients $n$ in the sum (\ref{N23})
contain sine functions, which are zero in the middle of their
period. Therefore, for $|\alpha | = T_s/2$, the odd $n$ terms
almost disappear and we get approximately an even coherent state.
We analyze in detail the $W$ functions for even $s$ only.
Nonetheless, for completeness of our discussion, we give the
explicit approximate expression for the quasiperiod
%----------------------------------------------------------------------
\begin{eqnarray}
 T_s\approx 2\sqrt{4s+6} \label{N35}
\end{eqnarray}
for odd $s$, which is twice larger than the quasiperiod given by
(\ref{N34}) for a chosen value of $s$.
%%%%%%%%%%%%%%%%%%%%%%%%%%%%%%%%%%%%%%%%%%%%%%%%%%%%%%%%%%%%%%%%%%%%%%%
% figure 4

%%%%%%%%%%%%%%%%%%%%%%%%%%%%%%%%%%%%%%%%%%%%%%%%%%%%%%%%%%%%%%%%%%%%%%%%
\subsection{B.~~ Truncated Coherent States}
\label{sectTCS} \inxx{coherent states; truncated}

Kuang et al. \cite{Kua93} defined the normalized FD coherent
states by truncating the Fock expansion of the conventional ID
coherent states or equivalently by the action of the operator
$\exp(\balpha{\hat a}^{\dagger})$ (with proper normalization) on
vacuum state. The Kuang et al. approach is similar to the
Vaccaro--Pegg treatment \cite{Vac90} of the Wigner function for
CS. The state $|\balpha\Rangle{s} $, where $\balpha =
|\balpha|\exp(\I \varphi)$, can be defined by its Fock expansion
\cite{Kua93}
%----------------------------------------------------------------------
\begin{eqnarray}
|\balpha\Rangle{s}  = \norm  \exp(\balpha{\hat a}_s^{\dagger})
\,|0\rangle = \sum_{n=0}^{s}b^{(s)}_n\,|n\rangle \label{N36}
\end{eqnarray}
with the Poissonian superposition coefficients
%----------------------------------------------------------------------
\begin{eqnarray}
b^{(s)}_n  =  \norm
\sum_{n=0}^{s}\frac{\overline\alpha^n}{\sqrt{n!}} \label{N37}
\end{eqnarray}
% figure 5
\begin{figure}
\vspace*{-4.3cm} \hspace*{3mm}
\centerline{\psfig{figure=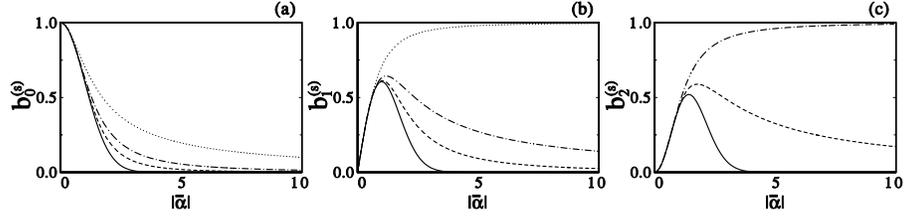,width=15cm}}
\vspace*{-15cm}  \caption{{\bf Truncated coherent states}:
Superposition coefficients ${b}^{(s)}_{n}$ of
$|\overline\alpha\rangle_{(s)}$ versus displacement parameter
amplitude $|\overline{\alpha}|$ for the same cases as in Fig.
\ref{mirafg02}.\label{mirafg05}}
%\vspace*{-1cm}
\end{figure}
%%%%%%%%%%%%%%%%%%%%%%%%%%%%%%%%%%%%%%%%%%%%%%%%%%%%%%%%%%%%%%%%%%%%%%%
% figure 6
\begin{figure}
\vspace*{-4.2cm} \hspace*{8mm}
\centerline{\psfig{figure=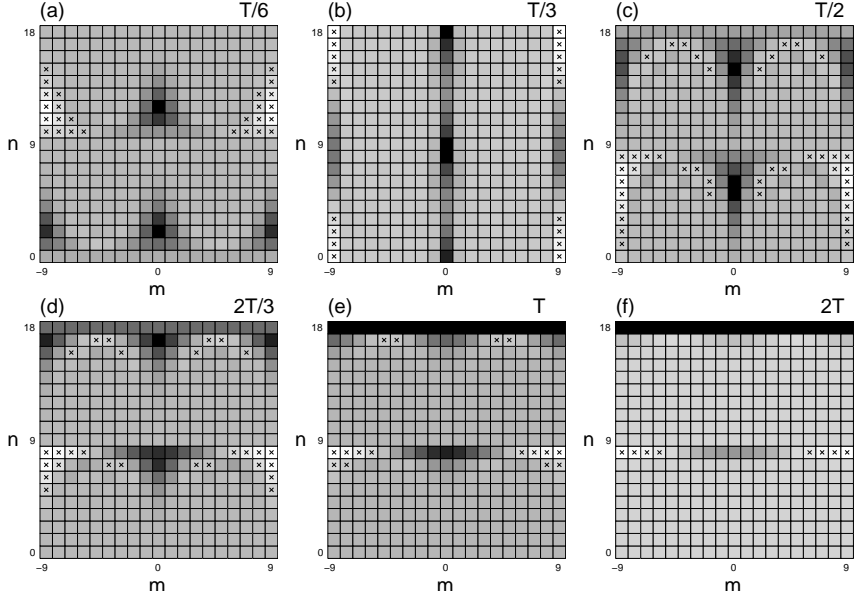,width=17cm}}
\vspace*{-12.5cm} \caption{{\bf Truncated coherent states}:
Wigner function for $|\overline{\alpha} \rangle_{(18)}$ with
different displacement parameters $\overline{\alpha}$ given by
fractions of $T=8.8$. \label{mirafg06}} \vspace*{-0.6cm}
\end{figure} \noindent%
normalized by
%----------------------------------------------------------------------
\begin{eqnarray}
\norm  &=& \Big(\sum_{n=0}^{s} \frac{|\balpha|^{2n}}{n!} \Big)
^{-1/2} \;=\; \left\{ (-1)^s {\rm L}^{-s-1}_s (|\balpha|^2)
\right\}^{-1/2} \label{N38}
\end{eqnarray}
where ${\rm L}^n_s(x)$ is the generalized Laguerre polynomial.
Equation (\ref{N36}) is just the Fock expansion of the
conventional ID CS, which are truncated at an $s$th term and
properly normalized. For this reason we shall refer to the state
(\ref{N36}) as the {\em truncated CS}. In Fig. \ref{mirafg05},
the superposition coefficients $b_{n}^{(s)}$, given by Eq.
(\ref{N37}) for the truncated CS $|\balpha \Rangle{s}$ are
presented as a function of the parameter
$|\balpha|\equiv|\alpha|$ in \Hilbert{s} with $s=1,2,3$ and
$s=\infty$. As seen in Fig. \ref{mirafg05}, the coefficients
$b^{(s)}_n$ are aperiodic functions of $|\balpha|$. We emphasize
the essential difference between the generalized and truncated
CS. The former are periodic or quasiperiodic, while the latter
are aperiodic in $|\balpha|=|\alpha|$. Nevertheless, both
$|\alpha\Rangle{s} $ and $|\balpha\Rangle{s} $, go over into the
conventional Glauber CS in the limit of $s\rightarrow\infty$ as is
convincingly depicted in Fig. \ref{mirafg03}. By definition, the
truncated CS go over into the Glauber CS in the limit of
$s\rightarrow\infty$. Nevertheless, for better comparison with the
generalized CS, given by (\ref{N20}), we show this property
explicitly by expanding the scalar products between
$|\balpha\Rangle{s} $ and $|\alpha\Rangle{\infty}$ in power series
of $|\alpha|$. We have \cite{Opa96}
%----------------------------------------------------------------------
\begin{eqnarray}
\Langle{\infty}\alpha|\balpha\Rangle{1} &=& 1 - \fra14 |\alpha|^4
+ \fra16 |\alpha|^6 - {\cal O}(|\alpha|^8)
\nonumber\\
\Langle{\infty}\alpha|\balpha\Rangle{2} &=& 1 -
\fra{1}{12}|\alpha|^6 + \fra{1}{16}|\alpha|^8 - {\cal
O}(|\alpha|^{10})
\nonumber\\
\Langle{\infty}\alpha|\balpha\Rangle{3} &=& 1 -
\fra{1}{48}|\alpha|^8 + \fra{1}{60}|\alpha|^{10} - {\cal
O}(|\alpha|^{12}) \label{N39}
\end{eqnarray}
where we put $\alpha=\balpha$. We find by induction that, with
increasing dimension $(s+1)$, the truncated CS approach the
conventional CS
%----------------------------------------------------------------------
\begin{eqnarray}
\Langle{\infty}\alpha|\balpha\Rangle{s}  &=& 1 -
\frac{|\alpha|^{2(s+1)}}{2\,(s+1)!} + {\cal O}(|\alpha|^{2(s+2)})
\label{N40}
\end{eqnarray}
for $|\alpha|\equiv|\balpha|^2\ll s$. Although Eqs. (\ref{N28})
and (\ref{N40}) have the same form, a closer comparison of Eqs.
(\ref{N27}) and (\ref{N39}) shows that the states
$|\alpha\Rangle{s} $ approach $|\alpha \Rangle{\infty}$ slower
than $|\balpha\Rangle{s} $ do and, in fact, the corrections ${\cal
O}(|\alpha|^{2(s+2)})$ in Eq. (\ref{N39}) are smaller than those
in Eq. (\ref{N27}). Finally, let us expand the scalar product
between $|\alpha\Rangle{s} $ and $|\balpha\Rangle{s} $ for $s$=1,
2, 3 in power series of $\alpha\equiv\balpha$. We find that
%----------------------------------------------------------------------
\begin{eqnarray}
\Langle{1}\alpha|\balpha\Rangle{1} &=& 1 - \fra{1}{18} |\alpha|^6
+ \fra{1}{15}|\alpha|^8 - {\cal
O}(|\alpha|^{10}) \nonumber\\
\Langle{2}\alpha|\balpha\Rangle{2} &=& 1 - \fra{1}{64}|\alpha|^8
+ \fra{9}{800} |\alpha|^{10} - {\cal
O}(|\alpha|^{12}) \nonumber\\
\Langle{3}\alpha|\balpha\Rangle{3} &=& 1 -
\fra{1}{300}|\alpha|^{10} + \fra{13}{5040} |\alpha|^{12} - {\cal
O}(|\alpha|^{14}) \label{N41}
\end{eqnarray}
The expansions up to $|\alpha|^{2(s+2)}$ can be written in
general form as
%----------------------------------------------------------------------
\begin{eqnarray}
 \Langle{s}\alpha|\balpha\Rangle{s}  &=&
1 - \frac{|\alpha|^{2(s+2)}}{2s!(s+2)^2} + {\cal
O}(|\alpha|^{2(s+3)}) \label{N42}
\end{eqnarray}
All these three types of CS are approximately equal for
$|\balpha|^{2}=|\alpha|^{2}\ll s$, since the scalar products
between them tend to unity. The higher $s$, the greater is the
range of $|\alpha|$, where the scalar product tends to unity.
However, the states are significantly different for values
$|\alpha|^{2}\approx s$. By comparing Eqs. (\ref{N28}),
(\ref{N40}) and (\ref{N42}) for the same $s$, we observe that
$|\alpha\Rangle{s} $ and $|\balpha\Rangle{s} $ approach each
other faster than $|\alpha\Rangle{\infty}$.

In order to calculate the Wigner function, we substitute Eq.
(\ref{N37}) into Eq. (\ref{N17}), arriving at
%----------------------------------------------------------------------
\begin{eqnarray}
W_s(n,\theta_{m}) &=& \frac{\norm^{2}}{s+1}\left(
\sum_{k=0}^{M}\frac{|\balpha|^{M}}{\sqrt{k!(M-k)!}} \exp[\I
(2k-M)(\theta_{m}-\varphi)]\right. \nonumber \\ &&\qquad+ \left.
\sum_{k=M+1}^{s}\frac{|\balpha|^{M+s+1}}{\sqrt{k!(M-k+s+1)!}}\right)
\label{N43}
\end{eqnarray}
where $M = 2n\ \mbox{mod}(s+1)$. Equation (\ref{N43}) can be
written in a form useful for a comparison with the Vaccaro--Pegg
result
%----------------------------------------------------------------------
\begin{equation}
 W_s(n, \theta _{m}) = \frac{1}{s+1}\Big[\Lambda _{1}(n,
|\balpha|)\Phi _{1}(n, \theta _{m}, \varphi )+ \Lambda _{2}(n,
|\balpha|) \Phi _{2}(n, \theta _{m}, \varphi )\Big] \label{N44}
\end{equation}
where
%----------------------------------------------------------------------
\begin{eqnarray}
 \Lambda _{1}(n, |\balpha|) &=& \frac{\norm^{2}|\balpha|^{M}}{\mu_1!}
\nonumber \\
\Phi _{1}(n, \theta _{m}, \varphi) &=& \mu_1!\sum_{k=0}^{M}
\frac{\cos [(2k-M)(\theta _{m}-\varphi )]}{[k!(M-k)!]^{1/2}}
\label{N45}
\end{eqnarray}
and
%----------------------------------------------------------------------
\begin{eqnarray}
\Lambda _{2}(n, |\balpha|) &=&
\frac{\norm^{2}|\balpha|^{M+s+1}}{\mu_2!}
\nonumber \\
\Phi _{2}(n, \theta _{m}, \varphi) &=& \mu_2!\sum_{k=M+1}^{s}
\frac{\cos [(2k-M-s-1)(\theta _{m}-\varphi )]}
{[k!(M-k+s+1)!]^{1/2}} \label{N46}
\end{eqnarray}
Here, $\mu_1=[\![M/2]\!]$ is the integer part of $M/2$, and
similarly $\mu_2=[\![(M+1+s)/2]\!]$. We note that the functions
$\Lambda_i$ do not depend on the phase $\varphi$ of $\balpha$ and
similarly the functions $\Phi_i$ do not depend on its amplitude
$|\balpha|$. In the Vaccaro--Pegg treatment, $|\balpha|^{2}$ was
always much less than $s$, so that the second term of Eq.
(\ref{N44}) could be neglected. Then the Wigner function was
factorizable into the amplitude dependent function $\Lambda _{1}$
and the phase dependent function $\Phi_1$, and the normalizing
constant $\norm$ was approximated by $\exp (-|\balpha|^{2}/2)$. It
can be seen that for general values of $\balpha$ of the truncated
CS this factorization is no longer feasible. Moreover, for large
$|\balpha|$, the second term of Eq. (\ref{N44}) becomes
predominant. We compare different shapes of the $W$ functions for
various $\balpha$ in Fig. \ref{mirafg06}. The functions are
computed for $s=18$.  With $|\alpha |$ increasing from zero, the
shape of the generalized CS is initially very similar to that of
the truncated CS (see Fig. \ref{mirafg04}). It occurs up to the
peak--antipeak transition from $n=s$ to $n=0$ around the value
$|\balpha|^{2} \approx s/2$ (corresponding to $|\alpha | \approx
T_{s}/3$ in Fig. \ref{mirafg06}b). However, if $|\balpha|^{2} \gg
s/2$, the situation is inverse: the second term of Eq.
(\ref{N44}) is now predominant and we observe two peaks for $n >
s/2$ and a peak --- antipeak structure for $n \le s/2$ (for
instance, Fig. \ref{mirafg06}d). In the case when $|\balpha|^{2}
\approx s$ (Fig. \ref{mirafg06}c), the $W$ function has a more
general shape. With increasing $|\balpha|$ the two-peak structure
shifts to larger values of $n$, while the peak--antipeak
structure gradually vanishes at $n \le s/2$ (Fig.
\ref{mirafg06}d--e). The shape is still comparatively simple
because the function Eq. (\ref{N44}) is a sum of only two
factorizable terms. By further increasing $|\balpha | \gg T_{s}$,
the Wigner function has a very simply structure representing the
number state $|s\rangle$. Even for $|\balpha | = 2T_{s}$, as
presented in Fig. \ref{mirafg06}f, the peak--antipeak structure
at $n \le s/2$ vanishes almost completely. In the limit of
$|\balpha |^2/s\rightarrow \infty$, the truncated CS approaches
the number state $|s\rangle$. This conclusion can also be deduced
from the behavior of the superposition coefficients $b^{(s)}_n$
in their dependence on $|\balpha|$ as depicted Fig.
\ref{mirafg05}. In contrast to the generalized CS, the behavior of
the truncated CS is aperiodic in $|\balpha|$.

%%%%%%%%%%%%%%%%%%%%%%%%%%%%%%%%%%%%%%%%%%%%%%%%%%%%%%%%%%%%%%%%%%%%%%%%
\subsection{C.~~ Example: Two-dimensional Coherent States}
\label{sect2DCS} \inxx{coherent states; two-dimensional}

The simplest nontrivial FD states are those spanned in a
two-dimensional system, namely, for $s = 1$. States in such a
system have intensively been studied by authors dealing with the
general problem of finite-dimensional quantum-optical states
\cite{Buz92,Kua93,Gan94}. Here, we would like to discuss this
problem from other points of view. Two-dimensional systems are
well known in various fields of physics, and we thus can apply the
results and concepts to describe our situation. Examples of
realizations of such a system can be given by the spin projection
of spin-$\frac{1}{2}$ particle, two-level atom or quantum dot.
Hence, the CS in \Hilbert{1} [see Eqs. (\ref{N48}) and
(\ref{N50})] can, in fact, be identified with the coherent
spin-$\frac{1}{2}$ state \cite{Rad71} or equivalently with the
two-level atomic coherent state \cite{Are72}. In the case of
$s$=1, the terms {\em photon number}, {\em phase} and {\em FD
harmonic oscillator} are a bit confusing and should be
understood, for example, as \cite{Woo87}: {\em $z$ component of
spin divided by $\hbar$}, {\em angle of orientation about the $z$
axis} and {\em spin}, respectively, or equivalently as atomic
quantities \cite{Are72,Buz92}. We use the notion {\em
two-dimensional} space or states to be consistent with our
general terminology applied in earlier sections. Although, we are
aware that this terminology might be misleading. In this section
we use a Poincar\'e sphere representation for the description of
the states discussed and their properties, like various operator
averages and squeezing degrees. Finally, we present the $W$
function for two-dimensional CS.

It is well known that states in a two-dimensional system can be
described by means of the Stokes parameters and visualized by
means of the Poincar\'e sphere. The density matrix of any
two-state system can be written in the form
%----------------------------------------------------------------------
\begin{eqnarray}
 \hat \rho &=& \frac{1}{2} \left(
 \begin{array}{cc}
 1+{\cal S}_{z} & {\cal S}_{x}+\I {\cal S}_{y} \\
 {\cal S}_{x}-\I {\cal S}_{y} & 1-{\cal S}_{z}
 \end{array} \right) \label{N47}
\end{eqnarray}
where ${\cal S}_{x},\ {\cal S}_{y}$, and ${\cal S}_{z}$ are the
\inx{Stokes parameters}. Using these parameters as coordinates of
a point in three-dimensional space, any state corresponds to a
point on a unit radius sphere, the so-called Poincar\'e sphere.
Pure states are represented by points on the surface, while
mixed-state points lie inside the sphere. Now, using this tool,
we can display both the two-dimensional generalized and truncated
CS and compare their expressions.

For the two-dimensional generalized CS, given by
%--------------------------------------------------------------------------
\begin{eqnarray}
\,|\alpha\Rangle{1} = \cos |\alpha| |0\rangle + \exp (\I \varphi
)\sin |\alpha |\,|1\rangle \label{N48}
\end{eqnarray}
the Stokes parameters are found to be
%----------------------------------------------------------------------
\begin{eqnarray}
 {\cal S}_{x} &=& \sin 2|\alpha | \cos \varphi
\nonumber \\
{\cal S}_{y} &=& -\sin 2|\alpha | \sin \varphi
\nonumber \\
{\cal S}_{z} &=& \cos 2|\alpha | \label{N49}
\end{eqnarray}
We note that any pure state in \Hilbert{1} is coherent. The
interpretation of the parameter $\alpha$ is very simple; its
module is proportional to the polar coordinate, while its
argument $\varphi$ is the azimuthal coordinate of the
representative Poincar\'e sphere point.

Similarly, we find the Stokes parameters for the truncated CS,
given by (\ref{N36}). The two-dimensional state
$|\balpha\Rangle{1}$, with the parameter $\balpha = |\balpha|\exp
(\I \varphi)$, is expressed by
%----------------------------------------------------------------------
\begin{eqnarray}
|\balpha\Rangle{1} &=& \frac{1}{\sqrt{1+|\balpha|^2}}\,|0\rangle+
\exp (\I \varphi )
\frac{|\balpha|}{\sqrt{1+|\balpha|^2}}\,|1\rangle
\nonumber\\
&=& \cos (\arctan|\balpha|) |0\rangle + \exp (\I \varphi ) \sin
(\arctan|\balpha|)\,|1\rangle \label{N50}
\end{eqnarray}
The Stokes parameters are now
%----------------------------------------------------------------------
\begin{eqnarray}
 {\cal S}_{x} &=&
2\frac{|\balpha|}{1+|\balpha|^{2}} \cos \varphi
\nonumber \\
{\cal S}_{y} &=& -2\frac{|\balpha|}{1+|\balpha|^{2}} \sin \varphi
\nonumber \\
{\cal S}_{z} &=& \frac{1-|\balpha|^{2}}{1+|\balpha|^{2}}
\label{N51}
\end{eqnarray}
The function of the argument $\varphi $ is the same as for the
generalized CS, while the meaning of the module $|\balpha|$ is
different from that of $|\alpha |$. We observe, for instance,
neither periodicity nor quasiperiodicity in $|\balpha|$. To
interpret $|\balpha|$, we write the last equation in (\ref{N51})
in the form $|\balpha|/(1-{\cal S}_{z}^{2})^{1/2} = 1/(1+{\cal
S}_{z})$. Thus, for a given $\balpha$, one can construct the
corresponding \inx{Poincar\'e sphere} point as follows
\cite{Opa96}: (1) by locating the complex number $\balpha$ in the
${\cal S}_{x}{\cal S}_{y}$ plane, so that the ${\cal S}_{x}$
($-{\cal S}_{y}$) coordinate is the real (imaginary) part of
$\balpha$, respectively; and (2) by connecting this point with the
lower pole of the Poincar\'e sphere by a straight line. The other
intersection of the line and the sphere is then the point
representing the coherent state.

For the case of two-dimensional CS, there have been computed
quantities such as the mean values and variances of the various
operators, including $\hat{N}$ and $\hat{\Phi}$ quadratures, and
their commutators \cite{Buz92,Kua93}. Most of these quantities
can easily be displayed on the Poincar\'e sphere and expressed by
means of the Stokes parameters.  We find that the following mean
values and variances are given respectively by
%----------------------------------------------------------------------
\begin{eqnarray}
 \langle{\hat N}\rangle &=& \frac{1 - {\cal S}_{z}}{2}
\nonumber \\
\langle(\Delta {\hat N})^2\rangle &=& \frac{{\cal
S}_{x}^{2}+{\cal S}_{y}^{2}}{4}
\nonumber \\
\langle{\hat \Phi}\rangle &=& \frac{(1 - {\cal S}_{x})\pi}{2}
\nonumber \\
\langle(\Delta {\hat \Phi})^2\rangle &=& \frac{{\cal S}_{y}^{2} +
{\cal S}_{z}^{2})\pi ^{2}}{4} \label{N52}
\end{eqnarray}
and the mean value of the $\hat N-\hat \Phi $ commutator is
%----------------------------------------------------------------------
\begin{eqnarray}
 \langle [\hat N, \hat \Phi ] \rangle &=&
\frac{\I   \pi {\cal S}_{y}}{2} \label{N53}
\end{eqnarray}
The degrees of squeezing $S_{N}$ and $S_{\Phi}$ are defined by
\inxx{squeezing; degree}
%----------------------------------------------------------------------
\begin{eqnarray}
 S_{N} &=& 2\langle(\Delta {\hat N})^2\rangle\,
|\langle [\hat N,\hat\Phi ] \rangle |^{-1}-1
\nonumber \\
S_{\Phi} &=& 2\langle(\Delta {\hat \Phi})^2\rangle\, |\langle
[\hat N,\hat\Phi ] \rangle |^{-1}-1 \label{N54}
\end{eqnarray}
They can be written in terms of the Stokes parameters as
%---------------------------------------------------------------------
\begin{eqnarray}
S_{N} &=& \frac{1}{\pi} \frac{{\cal S}_{x}^{2} + {\cal
S}_{y}^{2}} {|{\cal S}_{y}|} - 1
\nonumber \\
S_{\Phi} &=& \pi \frac{{\cal S}_{y}^{2} + {\cal S}_{z}^{2}}
{|{\cal S}_{y}|} - 1 \label{N55}
\end{eqnarray}
We found for the case of $s = 1$ that the averages of the
quantum-optical quantities are simply related to the Stokes
parameters. The correspondence can also be expressed in terms of
the operators $\hat N$ and $\hat \Phi $ in relation to the Pauli
matrices $\hat \sigma _{z}$ and $\hat \sigma _{x}$, or the
quadratures $\hat X_{a}$ and $\hat Y_{a}$ related to $\hat \sigma
_{x}$ and $\hat \sigma _{y}$.

Finally, we find the explicit expression for the Wigner function
in $n$ and $\theta$ for two-dimensional generalized CS. We get
%----------------------------------------------------------------------
\begin{eqnarray}
W_s(n, \theta _{m}) &=& \fra14\Bigl[1 + (-1)^{n}\cos (2|\alpha | )
\nonumber \\
&&+\; (-1)^{m}\sqrt{2} \sin (2|\alpha | ) \cos \Bigl(\varphi -
(-1)^{n} \fra{\pi}{4}\Bigr) \Bigr]\label{N56}
\end{eqnarray}
On simple replacement of $|\alpha|$ by $\arctan|\balpha|$ in Eq.
(\ref{N56}), one obtains the $W$ function for the two-dimensional
truncated CS.

%%%%%%%%%%%%%%%%%%%%%%%%%%%%%%%%%%%%%%%%%%%%%%%%%%%%%%%%%%%%%%%%%%%%%%%%
\section{V.~~ OTHER FD QUANTUM-OPTICAL STATES}
\label{sect5}

Analogously to the {\em generalized CS} in a FD Hilbert space,
analyzed in Section IV.A, other states of the electromagnetic
field can be defined by the action of the FD displacement or
squeeze operators. In particular, FD displaced phase states and
coherent phase states were discussed by Gangopadhyay \cite{Gan94}.
Generalized displaced number states and Schr\"odinger cats were
analyzed in Ref. \cite{Mir97} and generalized squeezed vacuum was
studied in Ref. \cite{Mir98}. A different approach to
construction of FD states can be based on truncation of the Fock
expansion of the well-known ID harmonic oscillator states. The
same construction, as for the {\em truncated CS}, was applied to
analyze, for instance, truncated Schr\"odinger cats by Zhu and
Kuang \cite{Zhu94,Kua96}, Miranowicz et al. \cite{Mir97}, and Roy
and Roy \cite{Roy98}; truncated phase CS by Kuang and Chen
\cite{Kua94};  truncated displaced number states by Miranowicz,
et al. \cite{Mir97}, or truncated squeezed vacuum by Miranowicz et
al. \cite{Mir98}.

%%%%%%%%%%%%%%%%%%%%%%%%%%%%%%%%%%%%%%%%%%%%%%%%%%%%%%%%%%%%%%%%%%%%%%%%
\subsection{A.~~ FD Phase Coherent States}
\label{sectPCS}

Here, we study two kinds of FD phase coherent states associated
with the Pegg--Barnett Hermitian optical phase formalism
\cite{Peg88}. First states, referred to as the {\em generalized
phase CS} or coherent phase states, are generated by the action of
the phase displacement operator. This definition of the phase CS
was applied by Gangopadhyay \cite{Gan94} in close analogy to
Glauber's idea of the conventional CS. The second definition of
phase CS is based on another phase ``displacement'' operator
formally designed by Kuang and Chen \cite{Kua94}. We shall refer
to these states as the {\em truncated phase CS} to stress its
similarity to the {\em truncated CS} described in Section IV.B.
We construct the phase CS explicitly and derive their discrete
Wigner representation. The FD phase CS are not only mathematical
structures. A framework for their physical interpretation is
provided by cavity quantum electrodynamics and atomic physics.

%%%%%%%%%%%%%%%%%%%%%%%%%%%%%%%%%%%%%%%%%%%%%%%%%%%%%%%%%%%%%%%%%%%%%%%%
\subsection*{\em  1. Generalized Phase CS}
\label{sectTPCS} \inxx{phase coherent states; generalized}

Gangopadhyay \cite{Gan94} has proposed a definition of the
generalized phase CS in formal analogy to the generalized CS,
defined by Eq. (\ref{N20}). The main idea is to choose a
preferred phase state $|\theta _{0}\rangle$, and then to
construct the phase creation ($\hat\phi_s^{\dagger}$) and phase
annihilation ($\hat\phi_s$) operators analogously to the
conventional (photon-number) creation and annihilation operators.
The phase CS are then constructed by replacing vacuum $|0\rangle$
by $|\theta _{0}\rangle$, and the operators $\hat a_s$ and $\hat
a_s^{\dagger}$ by $\hat\phi_s$ and $\hat\phi_s^{\dagger}$,
respectively, as given by Eq. (\ref{N11}). Thus, the generalized
phase CS is defined to be \cite{Gan94}
%----------------------------------------------------------------------
\begin{eqnarray}
|\beta,\theta_{0}\Rangle{s}  = \hat{D}_s(\beta,\theta_0)
\,|\theta_{0}\rangle  \label{N57}
\end{eqnarray}
by the action of the \inx{phase displacement operator}
%----------------------------------------------------------------------
\begin{eqnarray}
\hat{D}^{(s)}(\beta,\theta_0)= \exp[\beta\hat \phi_s^{\dagger}
-\beta^*\hat\phi_s] \label{N58}
\end{eqnarray}
on the preferred phase state $|\theta_{0}\rangle$.  By
generalizing the method described in Appendix, one can find the
following phase-state representation of the generalized phase CS
\cite{Mir95}
%----------------------------------------------------------------------
\begin{eqnarray}
|\beta,\theta_{0}\Rangle{s}  &=& \sum_{m=0}^{s} \e ^{\I
(\mu-m_0)\varphi}\, b_m^{(s)} |\theta_{m}\rangle \label{N59}
\end{eqnarray}
%%%%%%%%%%%%%%%%%%%%%%%%%%%%%%%%%%%%%%%%%%%%%%%%%%%%%%%%%%%%%%%%%%%%%%%
% figure 7
\begin{figure}
\vspace*{-4.2cm} \hspace*{8mm}
\centerline{\psfig{figure=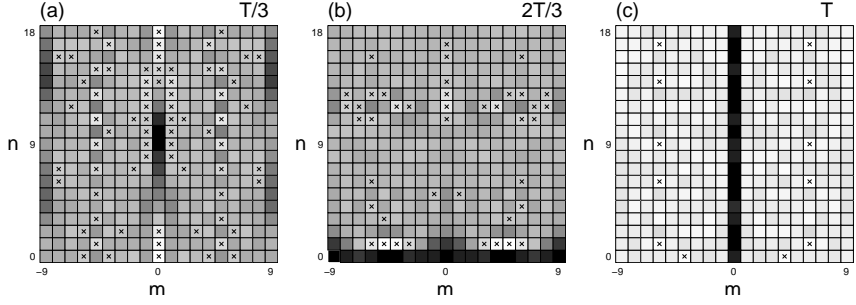,width=17cm}}
\vspace*{-16.5cm} \caption{{\bf Generalized phase coherent
states}: Wigner function for $|\beta,0\rangle_{(18)}$ with
different phase-displacement parameters $\beta$ chosen to be
fractions of the quasiperiod $T=T_{18}=8.8$. \label{mirafg07}}
\vspace*{-0.6cm}
\end{figure} \noindent %
where $\varphi={\rm Arg}\beta$ and the decomposition
coefficients are
%----------------------------------------------------------------------
\begin{eqnarray}
b_m^{(s)}  \;\equiv\; b_m^{(s)} (\theta_0) &=& \frac{s!}{s+1}
(-1)^{m+m_0} \, \frac{\I  ^{m_0+\mu}}{\sqrt{\mu! m_0!}}
\nonumber\\
&&\times\sum_{k=0}^{s} \exp\left(\I  x_k\gamma_s|\beta|\right)
\frac{{\rm He}_\mu(x_k) {\rm He}_{m_0}(x_k)}{ {\rm
He}_{s}^{2}(x_k)} \label{N60}
\end{eqnarray}
Here, $x_{l} \equiv x_{l}^{(s+1)}$ are the roots of the
\inx{Hermite polynomial}, $\mbox{He}_{s+1}(x_{l}) = 0$. For
brevity, we have denoted $\mu = m+m_0\ \mbox{mod}(s+1)$ and
$\gamma_s = \sqrt\frac{2\pi}{s+1}$. The values $\theta_m$ are
chosen $\mbox{mod}(2\pi)$. We also assume that the permitted
values of $\theta_{0}$ are not completely arbitrary but restricted
to $\frac{2\pi}{s+1}m_0\ \mbox{mod}(2\pi)$ (where
$m_0=0,1,\dots$). In a special case, for $\theta_0=0$ and $s=1$,
the phase CS reduce to the state $|\beta,\theta_0=0\Rangle{1}$
studied by Gangopadhyay \cite{Gan94}. Here, for simplicity, we
also consider the case of $\theta_0=0$.

%%%%%%%%%%%%%%%%%%%%%%%%%%%%%%%%%%%%%%%%%%%%%%%%%%%%%%%%%%%%%%%%%%%%%%%%
\subsection*{\em  2. Truncated Phase CS}
\label{sectGPCS} \inxx{phase coherent states; truncated}

Kuang and Chen \cite{Kua94} defined the FD phase CS, denoted as
$|\bbeta,\theta_0\Rangle{s}$, by the action of the FD operator
$\exp({\overline\beta}\hat\phi_s^{\dagger})$ on the phase state
$|\theta_0\rangle$.  The reference phase $\theta_0$ is chosen as
zero \cite{Kua94}. Therefore, on comparing the explicit
expressions for $\hat{a}_s$ and $\hat\phi_s$, it is clear that the
states $|\bbeta,\theta_0\Rangle{s} $ are in close analogy to the
truncated CS \cite{Opa96}. For this reason we shall refer to the
states $|\bbeta,\theta_0\Rangle{s} $ as the {\sl truncated phase
CS} in \Hilbert{s}.  For completeness, we present the phase-space
expansion with ${\overline\beta} = |{\overline\beta} |\exp(\I
\varphi)$ given by \cite{Kua94}
%----------------------------------------------------------------------
\begin{eqnarray}
 |{\overline\beta},\theta_0 \Rangle{s}  &=&
\norm  \exp({\overline\beta}\hat\phi_s^{\dagger})
|\theta_0\rangle \;=\;\sum_{m=0}^{s} \e ^{\I  m\varphi}
\,b_m^{(s)} |\theta_m\rangle \label{N61}
\end{eqnarray}
where
%----------------------------------------------------------------------
\begin{eqnarray}
 b_{m}^{(s)} = \norm
\frac{(\gamma_s|{\overline\beta}|)^{m}}{\sqrt{m!}},\qquad \norm =
\left(\sum_{n=0}^{s} \frac{(\gamma_s|{\overline\beta}|)^{2n}}{n!}
\right) ^{-1/2} \label{N62}
\end{eqnarray}
and $\gamma_s = \sqrt\frac{2\pi}{s+1}$ as in Eq. (\ref{N60}). In
particular, squeezing properties of the truncated phase CS were
analyzed by Kuang and Chen \cite{Kua94}. They have paid special
attention to the two-dimensional case.

Although many properties of the phase CS are known by now, for
their better understanding it is very useful to analyze graphs of
their quasidistributions. The discrete Wigner function, as defined
by Wootters \cite{Woo87} (see also Ref. \cite{Vac90}), takes the
following form for $s>1$
%----------------------------------------------------------------------
\begin{eqnarray}
 W_s(n,\theta_m) &=& \frac{1}{s+1}\sum_{p=0}^{s}
b^{(s)}_{m+p}\, b^{(s)}_{m-p} \exp\left[-2\I
p\left(\frac{2\pi}{s+1}n+\varphi\right) \right] \label{N63}
\end{eqnarray}
for the generalized phase CS with $b^{(s)}_{n}$ given by
(\ref{N60}) and for the truncated phase CS with superposition
coefficients (\ref{N62}). In Eq. (\ref{N63}), the subscripts
$m\pm p$ are assumed to be $\mbox{mod}(s+1)$.  One can obtain the
particularly simple Wigner function for $s=1$ \cite{Woo87}.

The generalized phase CS, $|\beta,\theta_0\Rangle{s}$, and
truncated phase CS, $|\bbeta,\theta_0\Rangle{s} $, are associated
with the Pegg--Barnett formalism of the Hermitian phase operator
${\hat\Phi_s}$. The operators ${\hat\Phi_s}$, $\hat\phi_s$, and
$\hat\phi_s^{\dagger}$ do not exist in the conventional ID Hilbert
space \Hilbert{\infty}. Thus the generalized and truncated phase
CS are properly defined {\em only} in \Hilbert{s} of finite
dimension. States $|\beta,\theta_0\Rangle{s} $ and
$|\bbeta,\theta_0\Rangle{s} $, similarly to $|\alpha\Rangle{s} $
and $|\overline{\alpha}\Rangle{s}$, approach each other for
$|\beta|^2=|{\overline\beta}|^2\ll s/\pi$ \cite{Opa96}. This can
be shown explicitly by calculating the scalar product between
generalized and truncated phase CS. We find
($\beta={\overline\beta}$)
%----------------------------------------------------------------------
\begin{eqnarray}
\Langle{s}\beta,\theta_0|{\overline\beta},\theta_0\Rangle{s} &=&
1 - \frac{(\sqrt{\pi}|\beta|)^{2(s+2)}}{2s!(s+2)^2} + {\cal
O}(|\beta|^{2(s+3)}) \label{N64}
\end{eqnarray}
For values $|\beta|^2=|{\overline\beta}|^2\approx s/\pi$ or
greater than $s/\pi$, the differences between
$|\beta,\theta_0\Rangle{s} $ and $|\bbeta,\theta_0\Rangle{s} $
become significant.

In Fig. \ref{mirafg07}, a few examples of the Wigner function for
$|\beta,0\Rangle{18}$ are presented for different values of the
phase displacement parameter $\beta$. Because of space limits in
this chapter, the corresponding figures for the truncated phase
CS are not presented. In Fig. \ref{mirafg07}c, we observe that
$|\beta,0\Rangle{s}$ is quasiperiodic in $\beta$. Closer analysis
of Eq. (\ref{N60}), in comparison to (\ref{N23}), shows that the
quasiperiod $T_s$ for $|\beta,0\Rangle{s}$ is the same as that
for the generalized coherent states. Thus, it is given by Eq.
(\ref{N34}) for even $s$ and Eq. (\ref{N35}) for odd $s$. Yet,
the evolution of the generalized phase CS is more complicated
than that for the generalized CS, as seen on comparing Figs.
\ref{mirafg07}a,b with the corresponding Figs. \ref{mirafg04}b,d.
As was discussed in Ref. \cite{Mir95}, the truncated phase CS are
aperiodic in ${\overline\beta}$ for any dimension.

%%%%%%%%%%%%%%%%%%%%%%%%%%%%%%%%%%%%%%%%%%%%%%%%%%%%%%%%%%%%%%%%%%%%%%%%
\subsection{B.~~ FD Displaced Number States}
\label{sectDNS}
%%%%%%%%%%%%%%%%%%%%%%%%%%%%%%%%%%%%%%%%%%%%%%%%%%%%%%%%%%%%%%%%%%%%%%%
% figure 8
\begin{figure}
\vspace*{-4.3cm} \hspace*{8mm}
\centerline{\psfig{figure=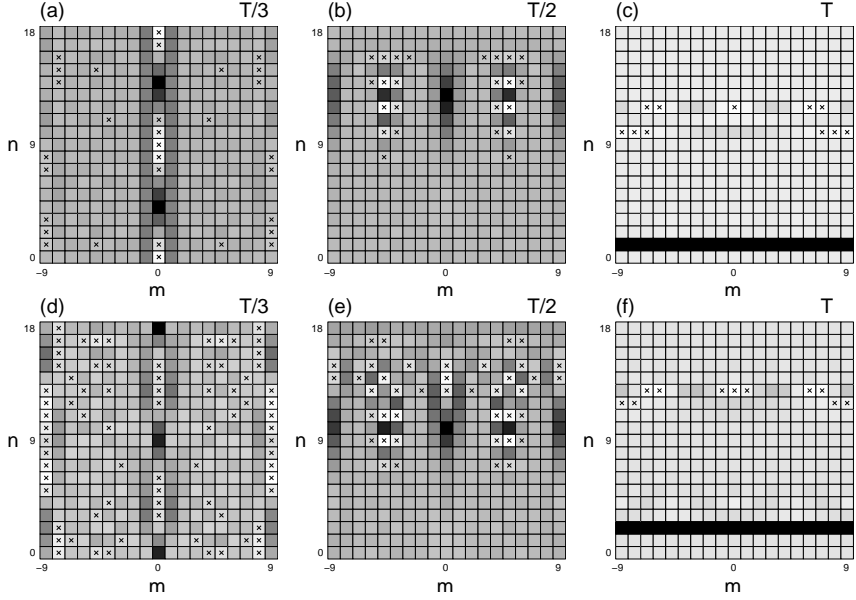,width=17cm}}
\vspace*{-12.5cm} \caption{{\bf Generalized displaced number
states}: Wigner function for $|\alpha,n_d\rangle_{(s)}=
|\alpha,1\rangle_{(18)}$ (a--c) and $|\alpha,2\rangle_{(18)}$
(d--f) with different displacement parameters $\alpha$ given by
fractions of the quasiperiod $T=T_{18}=8.8$. \label{mirafg08}}
\vspace*{-0.6cm}
\end{figure}

In this section, we propose two nonequivalent definitions of the
{\em displaced number states} (DNS) in the FD Hilbert space and
show that the FD states go over into the conventional DNS
discussed, for instance, by de Oliveira et al. \cite{Deo90}.

%%%%%%%%%%%%%%%%%%%%%%%%%%%%%%%%%%%%%%%%%%%%%%%%%%%%%%%%%%%%%%%%%%%%%%%%
\subsection*{\em  1. Generalized DNS}
\label{sectGDNS} \inxx{displaced number states; generalized}

Analogously to the generalized CS, given by Eq. (\ref{N20}), we
define the generalized DNS as follows
%----------------------------------------------------------------------
\begin{equation}\label{N65}
|\alpha, n_d\Rangle{s}={\hat D}^{(s)}(\alpha)|n_d\rangle
\end{equation}
as the result of action of the displacement operator $ {\hat
D}^{(s)}(\alpha)$, given by Eq. (\ref{N21}), on the number state
$|n_d\rangle$. By using the same method as described in Appendix
for \CS, we find the following explicit Fock representation of
the generalized DNS
%----------------------------------------------------------------------
\begin{equation}
|\alpha, n_d\Rangle{s} ={\hat D}_s(\alpha )|n_d\rangle
=\sum_{n=0}^s\e ^{\I (n-n_d)\varphi}\,b^{(s)}_n|n\rangle
\label{N66}
\end{equation}
where
%----------------------------------------------------------------------
\begin{equation}
 b^{(s)}_n\;\equiv \;b^{(s)}_n(n_d)=\frac{s!}{s+1}\
\frac{\,\I  ^{n_d}(-\I )^n}{\sqrt{n!n_d!}}\sum_{k=0}^s\e ^{\I
x_k|\alpha|}\,\frac{ {\rm He}_n(x_k){\rm He}_{n_d}(x_k)}{{\rm
He}_s^2(x_k)} \label{N67}
\end{equation}
and $\alpha =|\alpha|{\exp (\I \varphi )}$. In the dimension
limit, $s\rightarrow\infty$, the generalized DNS go over into the
conventional DNS defined, for instance, in Refs.
\cite{Deo90,Tan96}. This property can readily be deduced from
$\lim_{s\rightarrow \infty}\hat{D}_s=\hat{D}_{\infty}\equiv
\hat{D}$. Obviously, in the special case of $n_d=0$ the
generalized DNS reduce to the generalized CS defined by Eq.
(\ref{N20}).

%%%%%%%%%%%%%%%%%%%%%%%%%%%%%%%%%%%%%%%%%%%%%%%%%%%%%%%%%%%%%%%%%%%%%%%%
\subsection*{\em  2. Truncated DNS}
\label{sectTDNS} \inxx{displaced number states; truncated}

Let us define the finite-dimensional DNS, which in a special case
go over into the truncated CS of Kuang et al. \cite{Kua93} and
into the conventional DNS \cite{Deo90} in the limit of
$s\rightarrow \infty $. We define the  truncated displaced number
states, $|{ \overline{\alpha }},n_d\Rangle{s}$, by the following
Fock representation
%----------------------------------------------------------------------
\begin{equation}
|{\overline{\alpha },n_d\rangle }\s =\sum_{n=0}^sb^{(s)}_n\e ^{\I
(n-n_d)\varphi }|n\rangle  \label{N68}
\end{equation}
where
\begin{equation}
b^{(s)}_n\;\equiv \;b^{(s)}_n(n_d)=\norm
\left(\frac{n_1!}{n_2!}\right) ^{1/2}\,(-1)^{n_2-n}\,|\balpha
|^{n_2-n_1}L_{n_1}^{n_2-n_1}(|\balpha|^2) \label{N69}
\end{equation}
%----------------------------------------------------------------------
\begin{equation}
\norm \equiv \norm (|\overline{\alpha
}|,n_d)=\left(\sum_{n=0}^s\frac{n_1!}{n_2!}\,|\overline{\alpha
}|^{2(n_2-n_1)}\left[ L_{n_1}^{n_2-n_1}(|\overline{\alpha }
|^2)\right] ^2\right) ^{-1/2} \label{N70}
\end{equation}
For brevity, we have introduced the indices $n_1\equiv
\min(n,n_d)$ and $n_2\equiv \max(n,n_d)$. The state (\ref{N68})
is, in fact, given by the Fock expansion of the conventional ID
DNS \cite{Deo90,Tan96}, which are truncated at the $(s+1)$th term
and properly normalized. This construction justifies our name for
Eq. (\ref{N68}). Alternatively, the states (\ref{N68}) can be
defined by the action of the FD factorized displacement operator,
$\exp (\balpha{\ \hat{a}_s}^{\dagger })\exp(-\balpha^{*}
{\hat{a}_s})$, on a number state $|n_d\rangle$:
%----------------------------------------------------------------------
\begin{equation}
|{\overline{\alpha },n_d\rangle }\s =\norm \exp (\overline{\alpha
}{\hat{a}_s}^{\dagger })\exp ({-\overline{\alpha }}^{*}
\hat{a}_s)|n_d\rangle \label{N71}
\end{equation}
This equation explicitly shows how the concept of the truncated
CS, given by (\ref{N36}), is generalized. The truncated DNS are
different from the generalized DNS, given by (\ref{N65}). The
differences are particularly distinct for values $|\balpha
|^2\equiv |\alpha |^2$ of the order $s$ or greater. However, for
$|{ \overline{\alpha }}|^2\equiv |\alpha |^2\ll s,$ the FD
displaced number states $|\alpha ,n_d\Rangle{s} $ and
$|\balpha,n_d\Rangle{s} $ approach each other.

In Fig. \ref{mirafg08}, we present a few examples of the Wigner
function for the generalized DNS with $n_d$=1, and 2. We observe
that $|\alpha,n_d\rangle\s$ are quasiperiodic in $|\alpha|$ with
the same quasiperiods as those for the generalized CS given by
Eqs. (\ref{N34}) and (\ref{N35}) for even and odd $s$,
respectively. At multiples of $T_s$, the initial number state
$|n_d\rangle$ is partially recovered, as observed in Figs.
\ref{mirafg04}f and \ref{mirafg08}c,f. However, the periodicity
is deteriorated with increasing photon number $n_d$. The most
precise periodicity is observed for $|\alpha,0\rangle\s$, as
depicted in Fig. \ref{mirafg04}f. It is worse for
$|\alpha,1\rangle\s$ (Fig. \ref{mirafg08}c), and even worse for
$|\alpha,2\rangle\s$ as seen in Fig. \ref{mirafg08}f. In fact,
the entire evolution of $|\alpha,n_d\rangle\s$ becomes more
complicated with increasing number $n_d$ as can be observed by
comparing Figs. \ref{mirafg04}b,c,f, \ref{mirafg08}a--c and
\ref{mirafg08}d--f, respectively. For brevity, we omit the
corresponding figures for the truncated DNS. The Wigner functions
for $|\alpha ,n_d\Rangle{s} $ and $|\balpha,n_d\Rangle{s}$ are
almost indistinguishable for the displacement parameter
$|\alpha|=|\overline{\alpha}|$ and $n_d$ much less than $s$.
However, for higher values of these parameters, the generalized
and  truncated DNS behave qualitatively different. As discussed,
the former states are periodic or quasiperiodic, but the latter
are aperiodic with the increasing displacement parameter.

%%%%%%%%%%%%%%%%%%%%%%%%%%%%%%%%%%%%%%%%%%%%%%%%%%%%%%%%%%%%%%%%%%%%%%%%
\subsection{C.~~ FD Schr\"odinger Cats}
\label{sectSC} \inxx{even coherent states} \inxx{odd coherent
states}

Superpositions of two CS have attracted much attention
\cite{Mal79,Buz95} as simple examples of Schr\"odinger cats. In
this section, we will discuss two kinds of FD analogs of the
conventional ID even and odd CS of Malkin and Man'ko \cite{Mal79}.
The Schr\"odinger cats in FD Hilbert spaces were discussed, for
example, by  Zhu and Kuang \cite{Zhu94}, Miranowicz et al.
\cite{Mir97}, and Roy and Roy \cite{Roy98}.

%%%%%%%%%%%%%%%%%%%%%%%%%%%%%%%%%%%%%%%%%%%%%%%%%%%%%%%%%%%%%%%%%%%%%%%%
\subsection*{\em  1. Generalized Schr\"odinger Cats}
\label{sectGSC} \inxx{Schr\"odinger cats; generalized}
%%%%%%%%%%%%%%%%%%%%%%%%%%%%%%%%%%%%%%%%%%%%%%%%%%%%%%%%%%%%%%%%%%%%%%%
% figure 9
\begin{figure}
\vspace*{-4.2cm} \hspace*{8mm}
\centerline{\psfig{figure=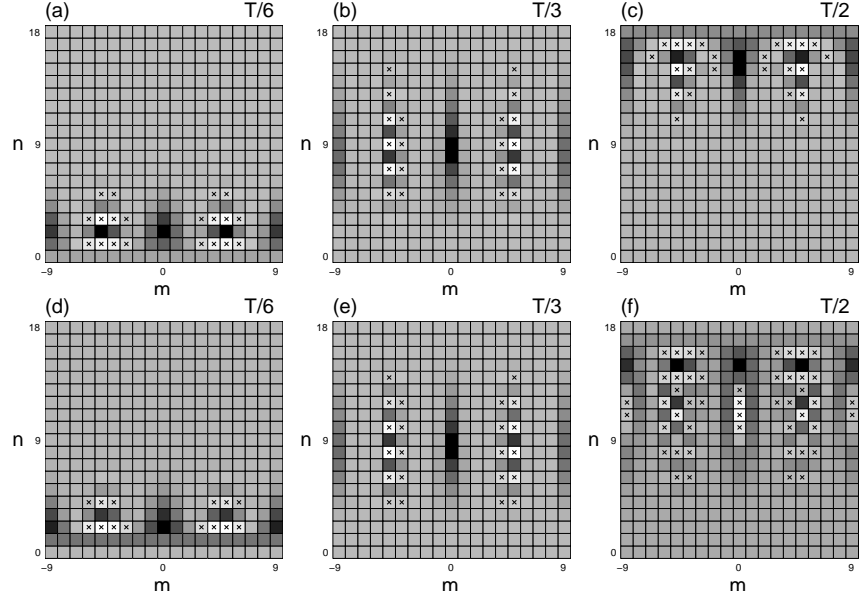,width=17cm}}
\vspace*{-12.5cm} \caption{{\bf Generalized Schr\"odinger cats}:
Wigner function for $|\alpha_0\rangle_{(18)}$ (a--c) and
$|\alpha_{1}\rangle_{(18)}$ (d--f) with $\alpha$ given by
fractions of the quasiperiod $T=T_{18}=8.8$. \label{mirafg09}}
\vspace*{-0.8cm}
\end{figure}

Let us define the generalized even CS by \cite{Mir97}
%----------------------------------------------------------------------
\begin{equation}
 |\alpha _{0}\Rangle{s} ={\cal
M}_{0s}\,\left(|\alpha \Rangle{s} +|-\alpha \Rangle{s} \right)
\label{N72}
\end{equation}
and odd CS by
%----------------------------------------------------------------------
\begin{equation}
 |\alpha _{1}\Rangle{s} ={\cal
M}_{1s}\,\left(|\alpha \Rangle{s} -|-\alpha \Rangle{s} \right)
\label{N73}
\end{equation}
where the normalization is guaranteed by ${\cal M}_{\delta s}$
($\delta=0,1$). On inserting Eq. (\ref{N22}) into (\ref{N72}) and
(\ref{N73}), we find the Fock expansions of the Schr\"odinger
cats in the forms
%----------------------------------------------------------------------
\begin{eqnarray}
 |\alpha _{0}\Rangle{s} &=& {\cal N}_{0s}
\sum_{n=0}^{[\![s/2]\!]} \e ^{\I 2n\varphi}\,b_{2n}^{(s)}|2n\rangle \nonumber\\
|\alpha _{1}\Rangle{s} &=&{\cal N}_{1s} \sum_{n=0}^{[\![s/2]\!]}
\e ^{\I (2n+1)\varphi}\,b_{2n+1}^{(s)}|2n+1\rangle \label{N74}
\end{eqnarray}
where the coefficients $b^{(s)}_n$ are given by Eq. (\ref{N23});
$[\![s/2]\!]$ is the integer part of $s/2$, and the normalizations
are ($\delta=0,1$)
%----------------------------------------------------------------------
\begin{equation}
{\cal N}_{\delta s}
=\left(\sum_{n=0}^{[\![s/2]\!]}\big(b^{(s)}_{2n+\delta})^2
\right)^{-1/2} \label{N75}
\end{equation}
Analogously one can construct FD superpositions of several CS,
that is FD Schr\"odinger cat-like or kitten states, which in the
limit go over into the conventional ID ones \cite{Buz95,Mir90}.

%%%%%%%%%%%%%%%%%%%%%%%%%%%%%%%%%%%%%%%%%%%%%%%%%%%%%%%%%%%%%%%%%%%%%%%%
\subsection*{\em  2. Truncated Schr\"odinger Cats}
\label{sectTSC} \inxx{Schr\"odinger cats; truncated}

FD even and odd CS can be constructed in a way slightly different
from that presented in the preceding paragraph. Instead of the
generalized CS, the truncated CS can be used in the definitions
(\ref{N72}) and (\ref{N73}). This approach was explored by Zhu
and Kuang \cite{Zhu94}, and Roy and Roy \cite{Roy98}. We rewrite
briefly their explicit expressions for $|\balpha_{0}\Rangle{s} $
and $|\balpha_{1}\Rangle{s} $ in Fock representation
%----------------------------------------------------------------------
\begin{equation}
|{\balpha_{\delta}\rangle }\s ={\cal N}_{\delta s}
\sum_{n=0}^{[\![s/2]\!]}\frac{{\balpha}^{2n+\delta }}{\sqrt{
(2n+\delta )!}}|2n+\delta \rangle \label{N76}
\end{equation}
where $\delta =0$ for even cats and $\delta =1$ for odd cats. The
normalization is
%----------------------------------------------------------------------
\begin{equation}
{\cal N}_{\delta s}=\left(\sum_{n=0}^{[\![s/2]\!]}\frac{|{
\balpha}|^{2(2n+\delta )}}{(2n+\delta )!}\right)
^{-1/2}\label{N77}
\end{equation}
Equation (\ref{N76}) can directly be calculated from Eqs.
(\ref{N72}) and (\ref{N73}) after replacing $|\pm
{\alpha}\Rangle{s}$  by $|\pm {\balpha} \Rangle{s}$, given by
their Fock expansion (\ref{N36}). Therefore, we refer to the
states (\ref{N76}) as the {\em truncated} states.

Several examples of the Wigner function for the generalized even
and odd CS are presented in Fig. \ref{mirafg09}. Their
interpretation is quite clear. These are two-peak structures with
many interference fringes. The fringes in the Wigner function are
typical of superposition states, and are not observed for a
mixture of states.  The main difference between the Wigner
functions presented in Figs. \ref{mirafg09}a--c and
\ref{mirafg09}d--f consists in a shift of the interference
fringes. Let us note that the generalized CS for $|\alpha|=T_s/2$,
presented in Fig. \ref{mirafg04}c, is approximately equal to the
even CS for the same value of $|\alpha|$. Unfortunately, because
of space limitations here, the corresponding Wigner functions for
the truncated cats are skipped. Let us mention that only for small
displacement parameter $|\balpha|^2=|\alpha|^2\ll s$, the
truncated and generalized cats have similar properties since
approximately holds $|\alpha _{0}\Rangle{s}\approx|\balpha
_{0}\Rangle{s}$ and $|\alpha
_{1}\Rangle{s}\approx|\balpha_{1}\Rangle{s}$. However, for higher
values of $|\alpha|^2$ (roughly estimated to be greater than
$T_s/3$), discrepancies between the generalized and truncated
Schr\"odinger cats become essential since they are defined in
terms of the CS $|\pm\alpha\Rangle{s}$ and
$|\pm\balpha\Rangle{s}$ exhibiting different properties for large
$|\alpha|^2$ as seen by comparing Figs. \ref{mirafg04}c--f and
\ref{mirafg06}c--f. These discrepancies result from periodic or
quasiperiodic behavior of the generalized states and aperiodic
behavior of the truncated states.

%%%%%%%%%%%%%%%%%%%%%%%%%%%%%%%%%%%%%%%%%%%%%%%%%%%%%%%%%%%%%%%%%%%%%%%%
\subsection{D.~~ FD Squeezed Vacuum}
\label{sectSV}

Here, we discuss two kinds of FD squeezed vacuum. We will present
explicit forms of these states, which reveal the differences and
similarities between them and the conventional IF squeezed vacuum
\cite{Lou87} or FD coherent state. We will show that our states
are properly normalized in \Hilbert{s} of arbitrary dimension and
go over into the conventional squeezed vacuum if the dimension is
much greater than the square of the squeeze parameter. Squeezing
and squeezed states in FD Hilbert spaces were analyzed, in
particular, by W\'odkiewicz et al. \cite{Wod87}, Figurny et al.
\cite{Fig93}, Wineland et al. \cite{Win94}, Bu\v{z}ek et al.
\cite{Buz92}, and Opatrn\'y et al. \cite{Opa96}. An FD analog of
the conventional squeezed vacuum was proposed by Miranowicz et
al. \cite{Mir98}.

%%%%%%%%%%%%%%%%%%%%%%%%%%%%%%%%%%%%%%%%%%%%%%%%%%%%%%%%%%%%%%%%%%%%%%%
% figure 10
\begin{figure}[ht]
\vspace*{-4.cm} \hspace*{8mm}
\centerline{\psfig{figure=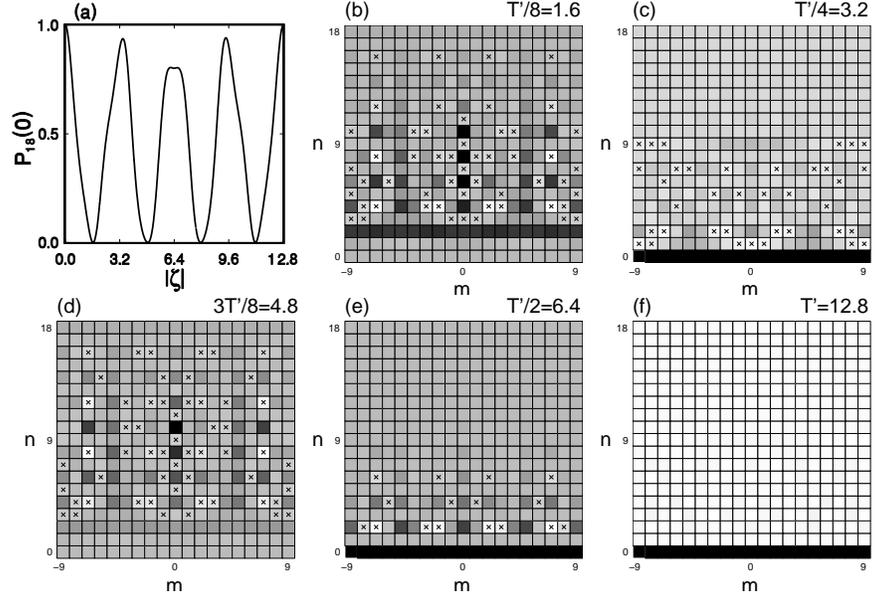,width=17cm}}
\vspace*{-12.5cm} \caption{{\bf Generalized squeezed vacuum}: (a)
Vacuum-state probability, $P_{18}(0)=|b_0^{(18)}|^2$, for
$|\zeta\rangle_{(18)}$ as a function of the squeeze parameter
amplitude $|\zeta|$; (b--f)  Wigner function for
$|\zeta\rangle_{(18)}$ with $\zeta$ given by fractions of the
quasiperiod $T'=T'_{18}=12.8$. \label{mirafg10}} \vspace*{-0.2cm}
\end{figure}

%%%%%%%%%%%%%%%%%%%%%%%%%%%%%%%%%%%%%%%%%%%%%%%%%%%%%%%%%%%%%%%%%%%%%%%%
\subsection*{\em  1. Generalized Squeezed Vacuum}
\label{sectGSV} \inxx{squeezed vacuum; generalized}

By analogy with the conventional squeezed vacuum \cite{Lou87}, we
define the generalized squeezed vacuum in the ($s+1$)-dimensional
Hilbert space by \cite{Mir98}
%----------------------------------------------------------------------
\begin{equation}
|\zeta \Rangle{s} =\hat{S}_s(\zeta )\,|0\rangle  \label{N78}
\end{equation}
as the result of action of the generalized FD \inx{squeeze
operator}
%----------------------------------------------------------------------
\begin{equation}
\hat{S}_s(\zeta )=\exp \left\{ \fra{1}{2}(\zeta
\hat{a}_{s}^{\dagger 2}-\zeta ^* \hat{a}_s^2)\right\} \label{N79}
\end{equation}
on vacuum. Here, $\zeta =|\zeta |\exp (\I \varphi )$ is the
complex squeeze parameter; $\hat{a}_s$ and $\hat{a}_s^{\dagger}$
are, respectively, the FD annihilation and creation operators
defined by Eq. (\ref{N08}). The method for finding explicit
number-state representation of the generalized CS, presented in
Appendix, can also be applied here. We find the following explicit
Fock expansion of the generalized squeezed vacuum
%----------------------------------------------------------------------
\begin{equation}
|\zeta \Rangle{s} =\sum\limits_{n=0}^\sigma b_{2n}^{(s)}\e ^{\I
n\varphi }|2n\rangle   \label{N80}
\end{equation}
with the superposition coefficients given by
%----------------------------------------------------------------------
\begin{equation}
b_{2n}^{(s)}=(-\I )^n\frac{(2\sigma
)!}{\sqrt{(2n)!}}\sum\limits_{k=0}^\sigma
\exp\left(\fra{\I}{2}|\zeta |x_k\right)\frac{G_n(x_k)} {G_\sigma
(x_k)G_{\sigma+1}^{^{\prime }}(x_k)}  \label{N81}
\end{equation}
where $\sigma =[\![s/2]\!]$ and $G_n(x)$ are the
\inx{Meixner--Sheffer orthogonal polynomials} defined by the
recurrence relation \cite{Mei34}
%----------------------------------------------------------------------
\begin{equation}
G_{n+1}=xG_n-2n(2n-1)G_{n-1}\quad\quad   \label{N82}
\end{equation}
for $n=2,3,\dots$, together with $G_0(x)=1$ and $G_1(x)=x$. In
Eq. (\ref{N81}), $x_k\equiv x_k^{(\sigma+1)}$ is the $k$th root ($
k=0,\dots,\sigma $) of the polynomial $G_{\sigma +1}(x)$ and
$G_{\sigma +1}^{\prime }({x_k})$ denotes the $x$-derivative at
${x=x_k}$. Since Eq. (\ref{N81}) is of a rather complicated form,
we present a few examples of the FD squeezed vacuum for small
dimensions. For $s$=2 and 3, we find
%----------------------------------------------------------------------
\begin{equation}
|\zeta \Rangle{2}=|\zeta \Rangle{3}=\cos \beta |0\rangle +\e ^{
\I \varphi }\sin \beta |2\rangle   \label{N83}
\end{equation}
where $\beta =\frac12 |\zeta  x_{0}^{(2)}| =\frac 1{\sqrt{2}
}|\zeta |$. For $s=4,5$, we have
%----------------------------------------------------------------------
\begin{equation}
|\zeta \rangle \!_{(4)}=|\zeta \rangle \!_{(5)}=\fra 17(6+\cos
\beta )|0\rangle +\;\e ^{\I \varphi }\fra 1{\sqrt{7}}\sin \beta
|2\rangle +\;\e  ^{2\I \varphi }\fra{2}{7}\sqrt{6}\sin^2(\fra 12
\beta) |4\rangle   \label{N84}
\end{equation}
where $\beta = \frac 12|\zeta x_{0}^{(3)}|=\sqrt{\frac 72}
|\zeta|$.  The generalized squeezed vacuum, given by (\ref{N80}),
has more complicated form in Fock basis than that for the
generalized CS, described by (\ref{N22}). In particular, the
solution (\ref{N81}) contains rather complicated Meixner--Sheffer
polynomials instead of the well-known Hermite polynomials, which
occur in the expansions for the generalized CS.

Here, we discuss only a few basic properties of generalized
squeezed vacuum, given by (\ref{N78}). By definition,  it is
properly normalized for arbitrary dimension of the Hilbert space.
There are several ways to prove that the generalized squeezed
vacuum  goes over into the conventional squeezed vacuum ($|\zeta
\rangle $) in the limit of $s\rightarrow \infty $. By definition
(\ref{N78}), one can conclude that the property
$\lim_{s\rightarrow \infty }|\zeta \Rangle{s}=|\zeta
\Rangle{\infty} =|\zeta \rangle $ holds, since the FD
annihilation and creation operators go over into the conventional
ones: $\lim_{s\rightarrow \infty }\hat{a}_s=\hat{a}$ and
$\lim_{s\rightarrow\infty} \hat{a}_s^{\dagger}=\hat{a}
^{\dagger}$. One can also show, at least numerically, that the
superposition coefficients (\ref{N81}) approach the coefficients
$b_n$ for the conventional squeezed vacuum: $\lim_{s\rightarrow
\infty }b^{(s)}_n=b_n$ for $n=0,\dots,s$. We apply another method
based on the calculation of the scalar product $\langle \zeta
|\zeta \Rangle{s} $. We show the analytical results for $|\zeta
|\leq 1$ only. We have found the scalar product between
conventional and generalized squeezed vacuums in the form (for
even $s$)
%----------------------------------------------------------------------
\begin{equation}
\langle \zeta |\zeta \Rangle{s} =\langle \zeta |\zeta \rangle
\!_{(s+1)}=1+\sum_{k=1}^\infty (-1)^k c_k^{(s)}|\zeta |^{s+2k}\leq
1 \label{N86}
\end{equation}
where the coefficients $c_k^{(s)}$ are positive and less than one
for any $k$ and $s$. We find that the explicit expansion up to
$|\zeta|^{s+2}$ can be given in terms of the binomial coefficient
as follows
%----------------------------------------------------------------------
\begin{equation}
\langle \zeta |\zeta \Rangle{s} = \langle \zeta |\zeta
\Rangle{s+1}= 1- \left({{s+1}}\atop{{\frac12 s+1}}\right)
\left(\frac{|\zeta|}{2}\right)^{s+2} +{\cal O}(|\zeta|^{s+4})
\label{N87}
\end{equation}
In particular, for $s=2,\dots,7$, we have
%----------------------------------------------------------------------
\begin{eqnarray}
\langle \zeta |\zeta \Rangle{2}=\langle \zeta |\zeta
\Rangle{3}&=&1-\frac 3{16}|\zeta |^4+\frac 18|\zeta |^6 -{\cal
O}(|\zeta|^{8})
\nonumber\\
\langle \zeta |\zeta \Rangle{4}=\langle \zeta |\zeta
\Rangle{5}&=&1-\frac 5{32}|\zeta |^6+\frac{185}{1024}|\zeta
|^8-{\cal O}(|\zeta|^{10})
\nonumber\\
\langle \zeta |\zeta \Rangle{6}=\langle \zeta |\zeta
\Rangle{7}&=&1-\frac {35}{256}|\zeta |^8+\frac{7}{30}|\zeta
|^{10}-{\cal O}(|\zeta|^{12}) \label{N88}
\end{eqnarray}
It is clearly seen that, for a given $\zeta $ , the scalar
products become closer to unity with increasing space dimension.
We conclude that the generalized state (\ref{N78}) approaches the
conventional squeezed vacuum in the dimension limit.

%%%%%%%%%%%%%%%%%%%%%%%%%%%%%%%%%%%%%%%%%%%%%%%%%%%%%%%%%%%%%%%%%%%%%%%%
\subsection*{\em 2. Truncated Squeezed Vacuum} \label{sectTSV}
\inxx{squeezed vacuum; truncated}

One can propose another definition of a FD squeezed vacuum, such
as by truncation of the Fock expansion of the conventional
squeezed vacuum at the state $|s\rangle$. Thus, we define the
truncated squeezed vacuum as follows \cite{Mir98}
%----------------------------------------------------------------------
\begin{equation}  \label{N89}
|\bzeta\Rangle{s}=\sum\limits_{n=0}^{\sigma +1}
{b}_{2n}^{(s)}\e^{\I n\varphi }|2n\rangle
\end{equation}
with the superposition coefficients
%----------------------------------------------------------------------
\begin{equation}  \label{N90}
{b}_{2n}^{(s)}=\norm  \frac{\sqrt{(2n)!}}{n!} {t}^n
\end{equation}
normalized by
%----------------------------------------------------------------------
\begin{equation}  \label{N91}
 \norm^{-2}=\cosh |\bzeta|-
2t^{2\sigma+2}\binom{2\sigma+1}{\sigma}\,{\rm _2F_1}\! (1,\left\{
\fra 32+\sigma ,2+\sigma\right\},4t^2)
\end{equation}
where $\sigma=[\![s/2]\!]$; $t=\frac 12\tanh |\bzeta|$, and ${\rm
_2F_1}$ is the generalized hypergeometric function. We marked with
a bar the complex squeeze parameter $\bzeta$ for the truncated
states in order to distinguish it from the generalized squeezed
vacuum defined by applying the FD squeeze operator. We give an
example of the truncated squeezed vacuum. For $s=2,3$, Eq.
(\ref{N89}) reduces to
%----------------------------------------------------------------------
\begin{equation}  \label{N92}
|\bzeta\Rangle{2}=|\bzeta\Rangle{3}=
\frac{|0\rangle+\sqrt{2}t|2\rangle}{\sqrt{1 + 2t^2}}
\end{equation}
The state (\ref{N89}), by definition, goes over into the
conventional squeezed vacuum in the limit of large dimension:
$\lim _{s\rightarrow \infty }|\bzeta \Rangle{s}=|\bzeta \rangle
\equiv |\zeta \rangle $. We can explicitly show this by expanding
the scalar products between them in power series with respect to
$|\zeta|^2=|\bzeta|^2\ll s$. We find (for even $s$)
%----------------------------------------------------------------------
\begin{equation}  \label{N93}
\Langle{\infty} \zeta|\bzeta\Rangle{s}= \Langle{\infty}
\zeta|\bzeta\Rangle{s+1}
\end{equation}
under assumption $\varphi =\overline{\varphi }$. In particular,
we have
%----------------------------------------------------------------------
\begin{eqnarray}  \label{N94}
\Langle{\infty} \zeta|\bzeta\Rangle{2}= \Langle{\infty}
\zeta|\bzeta\Rangle{3} &=& 1-\frac 3{16}|\zeta |^4+\frac
{3}{16}|\zeta |^6 -{\cal O}(|\zeta|^{8})
\nonumber\\
\langle \zeta |\bzeta \Rangle{4}=\langle \zeta |\bzeta
\Rangle{5}&=&1-\frac 5{32}|\zeta |^6+\frac{65}{256}|\zeta
|^8-{\cal O}(|\zeta|^{10})
\nonumber\\
\langle \zeta |\bzeta \Rangle{6}=\langle \zeta |\bzeta
\Rangle{7}&=&1-\frac {35}{256}|\zeta |^8+\frac{119}{384}|\zeta
|^{10}-{\cal O}(|\zeta|^{10})
\end{eqnarray}
We can also explicitly compare $|\bzeta ,2n_0\Rangle{s}$ with
$|\zeta ,2n_0\Rangle{s}$ with the help of their scalar products.
In particular, by putting $\varphi =\overline{\varphi}$, we have
%----------------------------------------------------------------------
\begin{eqnarray}  \label{N95}
\Langle{2}\zeta|\bzeta\Rangle{2} =
\Langle{2}\zeta|\bzeta\Rangle{3} &=&1-\frac 1{16}|\zeta |^6+\frac
{7}{80}|\zeta |^8 -{\cal O}(|\zeta|^{10})
\nonumber\\
\Langle{4}\zeta|\bzeta\Rangle{4} =
\Langle{5}\zeta|\bzeta\Rangle{5} &=& 1-\frac{75}{2^{10}} |\zeta
|^8 + \frac{131}{2^{10}}|\zeta |^{10} -{\cal O}(|\zeta|^{12})
\nonumber\\
\Langle{6}\zeta|\bzeta\Rangle{6} =
\Langle{7}\zeta|\bzeta\Rangle{7} &=&1-\frac {49}{640}|\zeta
|^{10}+\frac {189}{2^{10}}|\zeta |^{12} -{\cal O}(|\zeta|^{14})
\end{eqnarray}
By comparing Eq. (\ref{N95}) with Eqs. (\ref{N88}) and
(\ref{N94}), we can conclude that the differences between the
generalized and truncated squeezed vacuums are smaller than those
between them and the conventional squeezed vacuum. All these
states coincide in the high-dimension limit. In Fig.
\ref{mirafg10}b--f, we have presented the Wigner representation
of the generalized squeezed vacuum for those values of the squeeze
parameter $|\zeta|$, which correspond to maximum and minimum
values of the vacuum-state probability given by
$P_{s}(0)=|b^{(s)}_0|^2$ (see Fig. \ref{mirafg10}a). We find that
the generalized squeezed vacuum is quasiperiodic in $|\zeta|$.
Although, its quasiperiod $T'_s$ differs from $T_s$ for the
generalized CS, phase CS, displaced number states or
Schr\"odinger cats, as given by Eqs. (\ref{N34}) and (\ref{N35}).
The truncated squeezed vacuum is aperiodic in $|\bzeta|$,
similarly to other truncated quantum-optical states discussed in
Sections IV and V.

%%%%%%%%%%%%%%%%%%%%%%%%%%%%%%%%%%%%%%%%%%%%%%%%%%%%%%%%%%%%%%%%%%%%%%%%%%
\section{VI.~ CONCLUSION}

We have compared two approaches to define finite-dimensional (FD)
analogs of the conventional quantum-optical states of
infinite-dimensional Hilbert space. We have contrasted (1) the
generalized coherent states (CS), defined by the action of the
generalized FD displacement operator on vacuum, with (2) the
truncated CS, defined by the normalized truncated Fock expansion
of the conventional Glauber CS. We have shown both analytically
and graphically that these CS constructed in FD Hilbert spaces
exhibit essentially different behaviors; the generalized CS are
periodic (for $s$=1,2) or quasiperiodic (for higher $s<\infty$)
functions of the displacement parameter, whereas the truncated CS
are aperiodic for any $s$ (even for $s$=1). Both the generalized
and truncated CS go over into the conventional CS  in the
dimension limit. Nevertheless, the truncated CS approach the
conventional Glauber CS faster than the generalized CS do.
Besides, as the special case, we have compared in detail the
two-dimensional CS. We have analyzed other finite-dimensional
quantum-optical states. In particular, we have discussed: (1) FD
phase coherent states, (2) FD displaced number states; (3) FD
Schr\"odinger cats (including even and odd CS); and (4) FD
squeezed vacuums. We have confronted two essentially different
ways of defining states in FD Hilbert spaces.  We have constructed
explicitly all of these states generated by various
finite-dimensional displacement or squeeze operators using the
method developed in Ref. \cite{Mir94} for the generalized CS. We
have also presented graphical representations of the discrete
number-phase Wigner function, which enabled us a very intuitive
understanding of the properties of the generalized and truncated
quantum-optical states.

%%%%%%%%%%%%%%%%%%%%%%%%%%%%%%%%%%%%%%%%%%%%%%%%%%%%%%%%%%%%%%%%%%%%%%%%%%
\section{APPENDIX}
\def\theequation{A.\arabic{equation}}
\setcounter{equation}{0}

Here, after Ref. \cite{Mir94}, we present a method for finding the
coefficients $b^{(s)}_{n}$ of Fock representation of the
generalized CS, given by Eq. (\ref{N22}).

The Baker-Hausdorf formula cannot be used to solve this problem
because the commutator of the annihilation $\hat{a}_s$ and
creation $\hat{a}_s^{\dagger}$ operators is not a $c$-number. A
numerical procedure, leading to the coefficients $b^{(s)}_{n}$,
was proposed by Bu\v{z}ek et al. \cite{Buz92}. In order to solve
this problem analytically \cite{Mir94}, it is of advantage to
express the conventional coherent state, $|\alpha\rangle$, in the
Fock representation in a different manner
%----------------------------------------------------------------------
\begin{eqnarray}
|\alpha\rangle &=& \sum_{n=0}^{\infty} \frac{(\alpha
a^{\dagger}-\alpha^*a)^n}{n!}|0\rangle \nonumber\\ &=&
\sum_{n=0}^{\infty}\sum_{k=0}^{[\![n/2]\!]}
\frac{\sqrt{(n-2k)!}}{n!}\ d_{n,n-2k} (-\alpha^*)^k\alpha^{n-k}
|n-2k\rangle \label{N96}
\end{eqnarray}
where
%----------------------------------------------------------------------
\begin{eqnarray}
d_{n,k}   &=& \left({ n \atop k}\right) (n-k-1)!! \label{N97}
\end{eqnarray}
and $[\![x]\!]$ is the integer part of $x$. Thus, Eq. (\ref{N96})
for the generalized CS can be rewritten as
%----------------------------------------------------------------------
\begin{eqnarray}
|\alpha\Rangle{s}  = \sum_{k=0}^{s}\left[\sum_{n=k}^{\infty}
\frac{\sqrt{k!}}{n!}\ d^{(s)}_{nk}
(-\alpha^*)^{(n-k)/2}\alpha^{(n+k)/2}\right] \,|k\rangle \equiv
\sum_{k=0}^{s}c^{(s)}_{k}|k\rangle \label{N98}
\end{eqnarray}
The problem reduces to derivation of the coefficients
$d^{(s)}_{n,k}$ satisfying the condition in the dimension limit
%----------------------------------------------------------------------
\begin{eqnarray}
\lim\limits_{s\rightarrow\infty} d_{nk}^{(s)} \;=\;
d_{n,k}^{(\infty)}&\equiv&
 d_{nk} \:=\: \left( n \atop k
\right)\, (n-k-1)!! \label{N99}
\end{eqnarray}
We obtain the following simple recurrence formula
%----------------------------------------------------------------------
\begin{eqnarray}
d_{nk}^{(s)} &=& \theta_k d_{n-1,k-1}^{(s)} +(k+1) \theta_{k+1}
d_{n-1,k+1}^{(s)} \label{N100}
\end{eqnarray}
with the conditions $d_{00}^{(s)}=1$ and $d_{n,n+k}^{(s)} = 0$
for $s,k>0$. In Eq. (\ref{N100}), $\theta_{n}$ is the
Heaviside function defined to be
%----------------------------------------------------------------------
\begin{eqnarray}
\theta_n\equiv \theta(s-n) =\left\{ {1 \quad\mbox{for}\quad s\geq
n} \atop {0 \quad\mbox{for}\quad s<n} \right. \label{N101}
\end{eqnarray}
The solution of the recurrence formula (\ref{N100}) is
%----------------------------------------------------------------------
\begin{eqnarray}
d_{nk}^{(s)} &=& \frac{s!}{k!(s+1)} \sum_{l=0}^{s} \frac{{\rm
He}_k(x_l)}{[{\rm He}_{s}(x_l)]^2} x_l^n \label{N102}
\end{eqnarray}
where $x_l\equiv x^{(s+1)}_{l}$ are the roots of the \inx{Hermite
polynomial} ${\rm He}_{s+1}(x)$. A solution similar  to ours
(\ref{N102}) was found by Figurny et al. \cite{Fig93} in their
analysis of the eigenvalues of the truncated quadrature
operators. On performing summation in Eq. (\ref{N98}) with the
coefficients $d_{nk}^{(s)}$, given by (\ref{N102}), one readily
arrives at
%----------------------------------------------------------------------
\begin{equation}
C_n^{(s)} = \frac{s!}{s+1} \frac1{\sqrt{n!}} \sum_{k=0}^{s}
\exp\left\{ \I [n(\varphi-\pi/2)+x_k \,|\alpha|] \right\}
\frac{{\rm He}_n(x_k)}{[{\rm He}_{s}(x_k)]^2}\equiv \e^{\I
n\varphi}b_n^{(s)} \label{N103}
\end{equation}
or, equivalently, Eq. (\ref{N23}). Our procedure provides the
coefficients $b^{(s)}_{n}$ in a closed analytical form. This is
the solution of the problem formulated by Bu\v{z}ek et al.
\cite{Buz92}.

%%%%%%%%%%%%%%%%%%%%%%%%%%%%%%%%%%%%%%%%%%%%%%%%%%%%%%%%%%%%%%%%%%%%%%%%%%
\subsection*{ACKNOWLEDGMENTS}

A. M. and W. L. thank J. Bajer, S. Dyrting, M. Koashi, T.
Opatrn\'y, \c{S}. K. \"Ozdemir, J. Pe\v{r}ina, K. Pi\c{a}tek, and
R. Tana\'s for their stimulating discussions. A. M. is indebted
to Prof. Nobuyuki Imoto for his hospitality at SOKEN.

%%%%%%%%%%%%%%%%%%%%%%%%%%%%%%%%%%%%%%%%%%%%%%%%%%%%%%%%%%%%%%%%%%%%%%%%%%

\printindex

\begin{references}

\parskip=0pt\itemsep=0pt

\bibitem{Wey31} H. Weyl, {\em Theory of Groups and Quantum
Mechanics}, Dover, New York, 1931.

\bibitem{Sch60} J. Schwinger, {\em Proc. Natl. Acad. Sci. (USA)}
{\bf 46}, 570 (1960); reprinted in {\em Quantum Kinematics and
Dynamics}, Benjamin, New York, 1970, p. 63.

\bibitem{San76} T. S. Santhanam and A. R. Tekumalla, {\em Found.
Phys.} {\bf 6}, 583 (1976); T. S. Santhanam, {\em Phys. Lett. A}
{\bf 56}, 345 (1976); T. S. Santhanam and K. B. Sinha, {\em Aust.
J. Phys.} {\bf 31}, 233 (1978); T. S. Santhanam, in B. Gruber and
S. Millmann (Eds.), {\em Symmetries in Science}, Plenum, New
York, 1980, p. 337.

\bibitem{Rad71}
J. M. Radcliffe, {\em J. Phys. A} {\bf 4}, 313 (1971).

\bibitem{Are72} C. Arecchi, E. Courtens, R. Gilmore and H. Thomas,
{\em Phys. Rev. A} {\bf 6}, 2211 (1972).

\bibitem{Gla76}
R. J. Glauber and F. Haake, {\em Phys. Rev. A} {\bf 13}, 357
(1976).

\bibitem{Sch26}
E. Schr\"odinger, {\em Naturwissenschaften} {\bf 14}, 664 (1926).

\bibitem{Gla63}
R. J. Glauber, {\em Phys. Rev.} {\bf 130}, 2529 (1963); ibid.
{\bf 131}, 2766 (1963).

\bibitem{Sud63}
E. C. G. Sudarshan, {\em Phys. Rev. Lett.} {\bf 10}, 277 (1963).

\bibitem{Per72}
A. M. Perelomov, {\em Commun. Math. Phys.} {\bf 26}, 222 (1972);
{\em Usp. Fiz. Nauk} {\bf 123}, 23 (1977)
[{\em Sov. Phys. Usp.} {\bf 20}, 703 (1977)];
{\em Generalized Coherent States and their Applications},
Springer, Berlin, 1986.

\bibitem{Gil72}
R. Gilmore, {\em Ann. Phys. (N.Y.)} {\bf 74}, 391 (1972); {\em
Rev. Mex. Fis.} {\bf 23}, 142 (1974); {\em J. Math. Phys.} {\bf
15}, 2090 (1974).

\bibitem{Zha90} W. M. Zhang, D. H. Feng, and R. Gilmore, {\em Rev. Mod.
Phys.} {\bf 62}, 867 (1990).

\bibitem{Mal79} I. A. Malkin and V. I. Man'ko, {\em Dynamical
Symmetries and Coherent States of Quantum Systems}, Nauka,
Moscow, 1979.

\bibitem{Kla85}
J. R. Klauder and B. S. Skagerstam (Eds.), {\em Coherent States:
Applications in Physics and Mathematical Physics}, World
Scientific, Singapore, 1985.

\bibitem{Fen94}
D. H. Feng and J. Klauder (Eds.), {\em Coherent States: Past, Present and
Future}, World Scientific, Singapore, 1994.

\bibitem{Buz92}
V. Bu\v{z}ek, A. D. Wilson-Gordon, P. L. Knight, and W. K. Lai,
{\em Phys. Rev. A} {\bf 45}, 8079 (1992).

\bibitem{Kua93}
L. M. Kuang, F. B. Wang, and Y. G. Zhou, {\em Phys. Lett. A} {\bf
183}, 1 (1993); {\em J. Mod. Opt.} {\bf 41}, 1307 (1994).
% truncated CS

\bibitem{Mir94}
A. Miranowicz, K. Pi\c{a}tek, and R. Tana\'s, {\em Phys. Rev. A}
{\bf 50}, 3423 (1994).

\bibitem{Pat95}
A. K. Pati and S. V. Lawande, {\em Phys. Rev. A} {\bf 51}, 5012
(1995).

\bibitem{Opa96}
T. Opatrn\'y, A. Miranowicz, and J. Bajer, {\em J. Mod. Opt.} {\bf
43}, 417 (1996).

\bibitem{Mir97}
A. Miranowicz, T. Opatrn\'y, and J. Bajer, in T. Hakio\v{g}lu and
A. S. Shumovsky (Eds.), {\em Quantum Optics and the Spectroscopy
of Solids: Concepts and Advances}, Vol. 83, {\em Fundamental
Theories of Physics}, Kluwer, Dordrecht, 1997, p. 225.

\bibitem{Roy97a} B. Roy and R. Roychoudhury,
% Even and odd q-coherent states in a finite-dimensional basis
% and their squeezing properties,
{\em Int. J. Theor. Phys.} {\bf36}, 1525 (1997).

\bibitem{Roy97b} B. Roy,
% Higher moment properties of k-component (k>=3) coherent states
% in a finite dimensional basis,
{\em Modern Phys. Lett. B} {\bf 11}, 963 (1997).

\bibitem{Roy98} B. Roy and P. Roy,
{\em J. Phys. A} {\bf 31}, 1307 (1998).
% Coherent states, even and odd coherent states in a finite-
% dimensional Hilbert space and their properties,

\bibitem{Zhu94}
J. Y. Zhu and L. M. Kuang, {\em Phys. Let. A} {\bf 193}, 227
(1994); {\em Chinese Phys. Lett. A} {\bf 11}, 424 (1994).
% truncated even and odd CS

\bibitem{Peg88}
D. T. Pegg, and S. M. Barnett, {\em Europhys. Lett.} {\bf 6}, 483
(1988); {\em Phys. Rev. A} {\bf 41}, 3427 (1989); S. M. Barnett
and D. T. Pegg, {\em J. Mod. Opt.} {\bf 36}, 7 (1989).

\bibitem{Kua94}
L. M. Kuang and X. Chen, {\em Phys. Rev. A} {\bf 50}, 4228 (1994);
{\em Phys. Lett. A} {\bf 186}, 8 (1994).
% truncated PCS

\bibitem{Gan94}
G. Gangopadhyay, {\em J. Mod. Opt.} {\bf 41}, 525 (1994).
% PCS

\bibitem{Mir95}
 A. Miranowicz, K. Pi\c{a}tek, T. Opatrn\'y, and R. Tana\'s,
{\em Acta Phys. Slov.} {\bf 45}, 391 (1995).

\bibitem{Roy97c} P. Roy and B. Roy,
% Remarks on the construction of a Hermitian phase operator,
{\em Quantum and Semiclas. Optics} {\bf 9}, L37 (1997).

\bibitem{Wod87}
K. W\'odkiewicz, P. L. Knight, S. J. Buckle, and S. M. Barnett,
{\em Phys. Rev. A} {\bf 35}, 2567 (1987).

\bibitem{Fig93}
P. Figurny, A. Or\l owski, and K. W\'odkiewicz, {\em Phys. Rev A}
{\bf 47}, 5151 (1993).

\bibitem{Win94}
D. J. Wineland, J. J. Bollinger, W. M. Itano, and D. J. Heinzen,
{\em Phys. Rev. A} {\bf 50}, 67 (1994).

\bibitem{Mir98}
A. Miranowicz, W. Leo\'nski, and R. Tana\'s, in D. Han et al.
(Eds.), NASA Conference Publication 206855, Greenbelt, MD, 1998,
p. 91.

\bibitem{Kua96} L. M. Kuang and J. Y. Zhu,
{\em J. Phys. A} {\bf 29}, 895 (1996).
% Even and odd PCS

\bibitem{state}
Special issue on quantum state preparation and measurement, {\em
J. Mod. Opt.} {\bf 44} (11/12) (1997).

\bibitem{Ulf95}  U. Leonhardt,
{\em Phys. Rev. Lett.} {\bf 74}, 4101 (1995); ibid. {\bf 76},
4293 (1996); {\em Phys. Rev. A} {\bf 53}, 2998 (1996).

\bibitem{Man97}
V. I. Man'ko, O. V.  Man'ko, {\em JETP} {\bf 85}, 430 (1997); V.
A. Andreev, V. I. Man'ko, {\em JETP} {\bf 87}, 239 (1998); V. I.
Man'ko, S. S. Safonov, {\em Phys. Atom. Nuclei} {\bf 61}, 585
(1998).

\bibitem{Buz97}
V. Bu\v{z}ek, G. Drobn\'y, G. Adam, R. Derka, P. L. Knight, {\em
J. Mod. Opt.} {\bf 44}, 2607 (1997); V. Bu\v{z}ek, R. Derka, G.
Adam, P. L. Knight, {\em Ann. Phys. (San Diego)} {\bf 266}, 454
(1998).

\bibitem{Wal96}
R. Walser, J. I. Cirac, P. Zoller,
{\em Phys. Rev. Lett.} {\bf 77}, 2658 (1996).

\bibitem{Ami98}
J. -P. Amiet and S. Weigert, {\em J. Phys. A} {\bf 31} L543
(1998); ibid. {\bf 32} L269 (1999); {\em J. Opt. B} {\bf 1} L5
(1999).

\bibitem{Wel99}
D. -G. Welsch, W. Vogel, and T. Opatrn\'y, in E. Wolf (Ed.), {\em
Progress in Optics}, Vol. 39, North-Holland, Amsterdam, 1999, p.
63.

\bibitem{Tan96}
R. Tana\'s, A. Miranowicz, and T. Gantsog, in E. Wolf (Ed.), {\em
Progress in Optics}, Vol. 35, North-Holland, Amsterdam, 1996, p.
355.

\bibitem{Peg98}
D. T. Pegg, L. S. Phillips, and S. M. Barnett, {\em Phys. Rev.
Lett.} {\bf 81}, 1604 (1998); S. M. Barnett and D. Pegg, {\em
Phys. Rev. A} {\bf 60}, 4965 (1999).

\bibitem{Kon00}
M. Koniorczyk, Z. Kurucz, A. G\'abris, and J. Janszky, {\em Phys.
Rev. A} {\bf 62}, 013802 (2000).

\bibitem{Par00}
M. G. A. Paris, {\em Phys.Rev. A} {\bf 62}, 033813 (2000).

\bibitem{Mir00}
A. Miranowicz, \c{S}. K. \"Ozdemir, N. Imoto, and M. Koashi, {\em
Mtg. Abstr. Phys. Soc. Jpn.} {\bf 62}, 108 (2000).

\bibitem{Leo94}
W. Leo\'nski and R. Tana\'s, {\em Phys. Rev. A} {\bf 49}, R20
(1994); W. Leo\'nski, {\em Phys. Rev. A} {\bf 54}, 3369 (1996);
W. Leo\'nski, S. Dyrting, and R. Tana\'s, {\em J. Mod. Opt.} {\bf
44}, 2105 (1997).

\bibitem{Leo97}
W. Leo\'nski, {\em Phys. Rev. A} {\bf 55}, 3874 (1997).

\bibitem{Mir96a}
A. Miranowicz, W. Leo\'nski, S. Dyrting, and R. Tana\'s, {\em Acta
Phys. Slov.} {\bf 46}, 451 (1996).

\bibitem{Leo01}
W. Leo\'nski and A. Miranowicz, ``Quantum-optical states in
finite-dimensional Hilbert space. II.~State generation'', Chapter
4, this volume.

\bibitem{Wig32}
E. P. Wigner, {\em Phys. Rev.} {\bf 40}, 749 (1932); for a review,
see M. Hillery, R. F. O'Connell, M. O. Scully, and E. P. Wigner,
{\em Phys. Rep.} {\bf 106}, 121 (1984); V. I. Tatarskii, {\em Sov.
Phys. Usp.} {\bf 26}, 311 (1983).

\bibitem{Str57}
R. L. Stratonovich, {\em Sov. Phys. JETP} {\bf 4}, 891 (1957); L.
Cohen and M. O. Scully, {\em Found. Phys.} {\bf 16}, 295 (1986).

\bibitem{Oco84}
R. F. O'Connell, and E. P. Wigner, {\em Phys. Rev. A} {\bf 30},
2613 (1984).

\bibitem{Woo87}
W. K. Wootters, {\em Ann. Phys.} {\bf 176}, 1 (1987).

\bibitem{Coh88} O. Cohendet, P. Combe, M. Sirugue, and M. Sirugue-Collin,
{\em J. Phys. A} {\bf 21}, 2875 (1988).

\bibitem{Vac90}
J. A. Vaccaro and D. T. Pegg, {\em Phys. Rev. A} {\bf 41}, 5156
(1990).

\bibitem{Opa95}
T. Opatrn\'y, V. Bu\v{z}ek, J. Bajer, and G. Drobn\'y, {\em Phys.
Rev. A} {\bf 52}, 2419 (1995).

\bibitem{Opa96a}
T. Opatrn\'y, D. -G. Welsch, and V. Bu\v{z}ek, {\em Phys. Rev. A}
{\bf 53}, 3822 (1996).

\bibitem{Hak98} T. Hakio\v{g}lu,
{\em J. Phys. A} {\bf 31}, 6975 (1998).
% Finite-dimensional Schwinger basis, deformed symmetries, Wigner
% function, and an algebraic approach to quantum phase

\bibitem{Luk93}
A. Luk\v s and V. Pe\v rinov\'a, {\em Phys. Scr. T} {\bf 48}, 94
(1993).

\bibitem{Lui98} A. Luis and J. Pe\v{r}ina,
{\em J. Phys. A} {\bf 31}, 1423 (1998).
% Discrete Wigner function for finite-dimensional systems

\bibitem{Per94}
J. Pe\v rina, Z. Hradil, and B. Jur\v co, {\em Quantum Optics and
Fundamentals of Physics}, Vol. 63, {\em Fundamental Theories in
Physics}, Kluwer Academic, Dordrecht, 1994.

\bibitem{Leo96}
W. Leo\'nski and A. Miranowicz, {\em Acta Phys. Slov.} {\bf 46},
433 (1996); W. Leo\'nski, A. Miranowicz, and R. Tana\'s, {\em
Laser Physics} {\bf 7}, 126 (1997).

\bibitem{Deo90}  F.~A.~M. de Oliveira, M. S. Kim, P. L. Knight, and V.
Bu\v{z}ek, {\em Phys. Rev. A} {\bf 41}, 2645 (1990).

\bibitem{Buz95} V. Bu\v{z}ek, and P. L. Knight, in E. Wolf (Ed.),
{\em Progress in Optics}, Vol. 34, North-Holland, Amsterdam,
1995, p. 1.

\bibitem{Mir90} A. Miranowicz, R. Tana\'s, and S. Kielich,
{\em Quantum Opt.} {\bf 2}, 253 (1990).

\bibitem{Lou87}
R. Loudon and P. L. Knight, {\em J. Mod.  Opt.} {\bf 34}, 709
(1987); K. Zaheer and M. S. Zubairy, in D. Bates and B. Bederson
(Eds.), {\em Advances in Atomic, Molecular and Optical Physics},
Vol. 28, Academic Press, New York, 1990, p. 143.

\bibitem{Mei34}
J. Meixner, {\em J. London Math. Soc.} {\bf 9}, 6 (1934); I. M.
Sheffer, {\em Duke Math. J.} {\bf 5}, 590 (1939).

\end{references}
\end{document}